\documentclass[apj]{emulateapj}
\usepackage{graphicx}
\usepackage{epstopdf}
\usepackage{soul}
\usepackage{amsmath}
\usepackage{natbib}
\usepackage{color}

\def\asec{$^{\prime\prime}$}
\def\lax{{$\mathrel{\hbox{\rlap{\hbox{\lower4pt\hbox{$\sim$}}}\hbox{$<$}}}$}}
\def\gax{{$\mathrel{\hbox{\rlap{\hbox{\lower4pt\hbox{$\sim$}}}\hbox{$>$}}}$}}

\slugcomment{Accepted for publication in {\it The Astrophysical Journal}}

\begin{document}

\title{On the limits of measuring the bulge and disk properties of local and
  high-redshift massive galaxies}
\author{Roozbeh Davari\altaffilmark{1,2},
Luis C. Ho\altaffilmark{3,4}, and
Chien Y. Peng\altaffilmark{5} 
}
\altaffiltext{1}{University of California, Riverside 900 University Avenue, Riverside, CA 92521, USA}
\altaffiltext{2}{The Observatories of the Carnegie Institution for Science 813 Santa Barbara Street, Pasadena, CA 91101, USA}
\altaffiltext{3}{Kavli Institute for Astronomy and Astrophysics, Peking University, Beijing 100871, P. R. China}
\altaffiltext{4}{Department of Astronomy, School of Physics, Peking University, Beijing 100871, P. R. China}
\altaffiltext{5}{Giant Magellan Telescope Organization 251 South Lake Avenue, Suite 300 Pasadena, CA 91101, USA }

\begin{abstract}

A considerable fraction of the massive quiescent galaxies at \emph{z}
$\approx$ 2, which are known to be much more compact than galaxies of
comparable mass today, appear to have a disk.  How well can we measure
the bulge and disk properties of these systems?  
We simulate two-component model galaxies in order 
to systematically quantify the
effects of non-homology in structures and
the methods employed. We employ empirical scaling relations to produce realistic-looking
 local galaxies with a uniform and wide range of bulge-to-total ratios
 ($B/T$), and then rescale them to mimic the signal-to-noise ratios
 and sizes of observed galaxies at \emph{z} $\approx$ 2.  This
 provides the most complete set of simulations to date for which we can examine the
 robustness of two-component decomposition of compact disk galaxies at
 different $B/T$. 
We confirm that the size of these massive, compact galaxies can be
measured robustly using a single S\'{e}rsic fit.
We can measure $B/T$ accurately without imposing any constraints on
the light profile shape of the bulge, but, due to the small angular
sizes of bulges at high redshift, their detailed properties can only
be recovered for galaxies with $B/T$ \gax\ 0.2.  The disk component,
by contrast, can be measured with little difficulty.   

\end{abstract}


\section{Introduction}

Discovery of compact, red massive galaxies at \emph{z} $\approx$ 2 (e.g.,
\citealt{Franx03}; \citealt{Daddi05}; \citealt{Kriek06}) have opened a new
door for improving the current models of galaxy formation and
evolution (e.g., \citealt{Wuyts10}; \citealt{Oser12};
\citealt{Ishibashi13}; \citealt{Dekel14}). Several studies have
confirmed the compactness of these galaxies (e.g., \citealt{Daddi05};
\citealt{Toft07}; \citealt{Trujillo07}; \citealt{Buitrago08};
\citealt{Cimatti08}; \citealt{Franx08}; \citealt{vanderWel08};
\citealt{vanDokkum08}; \citealt{Damjanov09}; \citealt{Hopkins09};
\citealt{Cassata10, Cassata11}; \citealt{Mancini10};
\citealt{Newman12}; \citealt{Szomoru12}; \citealt{Barro13};
\citealt{Huang13b}; \citealt{Williams14}). These ``red nuggets,'' while
common at \emph{z} $\approx$ 2, are rare in the local universe,
thus implies a considerable size increase  (3 -- 4 times) in the last 10 billion
years (\citealt{vanDokkum08};
\citealt{Trujillo09}; \citealt{Taylor10}; \citealt{vanDokkum10}; but
see \citealt{Saracco10}; \citealt{Valentinuzzi10}; \citealt{Ichikawa12};
\citealt{Poggianti13}). Red nuggets are found
to be as compact as $\sim 1-2$ kpc, which is comparable to the size of  the {\it
Hubble Space Telescope (HST)} point-spread function (PSF). This
raises the concern that the size and mass measurements of these galaxies
are subject to potential uncertainties (\citealt{Hopkins09}; \citealt{Muzzin09}). 

Besides the small size of the red nuggets, at least a considerable fraction of
them are found to have a disk, which imposes a further
challenge to our paradigm of galaxy evolution.
Van der Wel et al. (2011) claim that more than 50$\%$ of the
population of massive quiescent galaxies at \emph{z} $>$ 2 are
disk-dominated. \citet{Chang13}, by deprojecting the
observed axial ratios of the galaxies in their sample, show that
early-type galaxies at \emph{z} $>$ 1 are, on average, flatter than their counterparts at
\emph{z} $<$ 1. Furthermore, they claim that the median projected axis ratio at a fixed mass
decreases with redshift, which hints at the prevalence of disks at
higher redshifts. \citet{Patel13} find that their sample of quiescent massive
galaxies at higher redshifts have lower axial ratios ($b/a$) and concluded that the stars
in the progenitors of today's 2$M^{\star}$  galaxies were distributed in disks
at \emph{z} $\approx$ 3. \citet{Bruce14} show that within the redshift interval 1 $<$
\emph{z} $<$ 3 most massive galaxies are morphologically composite 
systems containing both a bulge and a disk component.  More than 80$\%$ of
their sample require a disk component to properly fit their light distribution.
  Aside from the observational evidence of
the prevalence of disks among the most massive galaxies at high redshift,
cosmological simulations predict the formation of massive disk
galaxies at these epochs. The compactness of red nuggets indicates
that these galaxies have experienced severe dissipation during their
formation. One plausible scenario is that the red nuggets are the end
result of major gas-rich mergers. \citet{Robertson06} find that
nearly all their simulated gas-rich merger remnants contain
rapidly rotating stellar substructure, while disk-dominated remnants
are restricted to form in mergers that are gas-dominated at the time
of final coalescence. They show that the formation of
rotationally supported stellar systems in mergers is not restricted to
idealized orbits, and both gas-rich major and minor mergers can
produce disk-dominated stellar remnants. Their findings can be
especially important for galaxy formation at high redshifts, where
gas-dominated mergers are common.  

Measuring the bulge and disk properties, and subsequently the luminosity 
bulge-to-total ratio ($B/T$), of galaxies obviously can reveal key properties 
and clues to formation and evolutionary paths that may be obscured by studying 
a bulge+disk galaxy as a single system.  Several factors can affect the reliability of bulge-disk
decomposition, including how the fitting pipeline is
employed, the galaxy brightness [signal-to-noise ratio ($S/N$)],
cosmological surface brightness dimming, and the effect of the
PSF. One of the best ways for quantifying the influence of these factors
is through galaxy simulations (e.g., \citealt{Trujillo07}, \citealt{Cimatti08},
\citealt{Mancini10}, \citealt{Szomoru10,Szomoru12},
\citealt{vanDokkum10}, \citealt{Williams10}, \citealt{Papovich12},
\citealt{vanderWel12}, and \citealt{Davari14} for high-\emph{z} galaxies,
and \citealt{Haussler07} and \citealt{Meert13} for low-\emph{z}
galaxies).

This work employs well-tested properties and scaling
relations of local galaxies to generate mock bulge+disk galaxies with a
uniform and wide range of $B/T$. This provides the most complete set of
simulations to date that allows us to examine the robustness of two-component
decomposition of disk galaxies at different $B/T$ values.  
Although our model bulge+disk galaxies do not capture the full observed complexity of local disk galaxies (e.g., \citealt{Gadotti09}; \citealt{KB10}), recent zoom-in cosmological simulations find that galaxies with bars and spiral structures are rare at $z \approx 2$ (\citealt{Kraljic12}).  If higher redshift bulges and disks resemble their local counterparts, the results of our rescaled model galaxies can be used as a yardstick for examining the robustness of these types of analysis for higher redshift galaxies. 
We address three key questions:

1) How well can single-component fitting of bulge+disk galaxies measure the global size and total luminosity of these galaxies?

2) Can we recover the properties of both the bulge and disk components, and if so, how well?

3) What are the best methods for measuring the $B/T$ of composite galaxies? And what are the potential biases of different bulge-disk fitting methods?

This paper is organized as follows. In Section 2, the details of the
galaxy simulations are provided. The main results are presented in
Section 3. Comparison with similar studies is done in Section
4. In Section 5, implications of our results for red nuggets are
discussed, and a summary is listed in Section 6. Results are based on
a standard cosmology ($\emph{H}_0$ = 71 ${\rm km^{-1}\, s^{-1}\,
 Mpc^{-1}}$, $\Omega_m$ = 0.27, and $\Omega_{\Lambda}$ = 0.73) and AB magnitudes.

\section{Method}

\subsection{Technique}

{\tt GALFIT 3.0} (\citealt{Peng10}) is utilized for our
simulations. {\tt GALFIT} is used extensively for modeling the light
profiles of galaxies. It provides
several commonly used functions in the astronomical literature. 
For our applications, we only use the S\'{e}rsic (1968) function to
model the surface brightness profiles: 

\begin{equation}
\Sigma(R) = \Sigma_e \ exp {\left \{-\kappa \left [ \left (\frac{R}{R_e} \right )^{1/n} - \  1\right ]\right \}},
\label{eq:sersic}
\end{equation}

\noindent where $R_e$ is the effective radius of the galaxy,
$\Sigma_e$ is the surface brightness at $R_e$, the S\'{e}rsic index
$n$ describes the profile shape, and the parameter $\kappa$ is closely
connected to $n$ (\citealt{Ciotti91}). The special cases of the S\'{e}rsic profile
are the exponential profile (\emph{n} = 1; \citealt{Freeman70}) and
the \emph{$R^{1/4}$} law (\emph{n} = 4; \citealt{deVaucouleur48}), which are
commonly observed in spiral and elliptical galaxies,
respectively. This suggests that the S\'{e}rsic index can be used as a
yardstick for distinguishing the disk-dominated from the
bulge-dominated galaxies (e.g., \citealt{Blanton03}; \citealt{Shen03};
\citealt{Bell04}; \citealt{Hogg04}; \citealt{Ravindranath04};
\citealt{Barden05}; \citealt{McIntosh05}; \citealt{Fisher08}).

\subsection{Simulated Model Galaxies}
 
We aim to simulate galaxies that resemble real, observed disk
galaxies. Toward this goal, we use empirical scaling relations and
other empirical constraints derived from observations of nearby
galaxies as inputs to create the model galaxies.  Although these
constraints reduce the generality of our simulated sample, they
provide us with simulated galaxies that exhibit realistic values of
bulge and disk component parameters. 

The well-known Kormendy relation (\citealt{Kormendy77};
\citealt{Hamabe87}), a projection of the galaxy fundamental plane
(\citealt{Djorgovski87}; \citealt{Dressler87}), 
reveals that elliptical galaxies and classical bulges follow
a correlation between effective surface brightness ($\mu_e$) and
effective radius ($R_e$).
The Kormendy relation indicates that larger elliptical
galaxies and classical bulges have lower densities. Since brighter
galaxies are bigger, a more general statement is that more luminous
systems are fluffier. The Kormendy relation, which has played an important role in 
the study of the formation and evolution of galaxies, has been studied
in different bands (e.g., \citealt{LaBarbera10}), environments (e.g.,
\citealt{Nigoche-Netro07}), redshifts (e.g., \citealt{LaBarbera03};
\citealt{Longhetti07}),  and magnitude ranges
(\citealt{Nigoche-Netro08}).   

\begin{figure*}[t]
  \centering
\includegraphics[width=180mm]{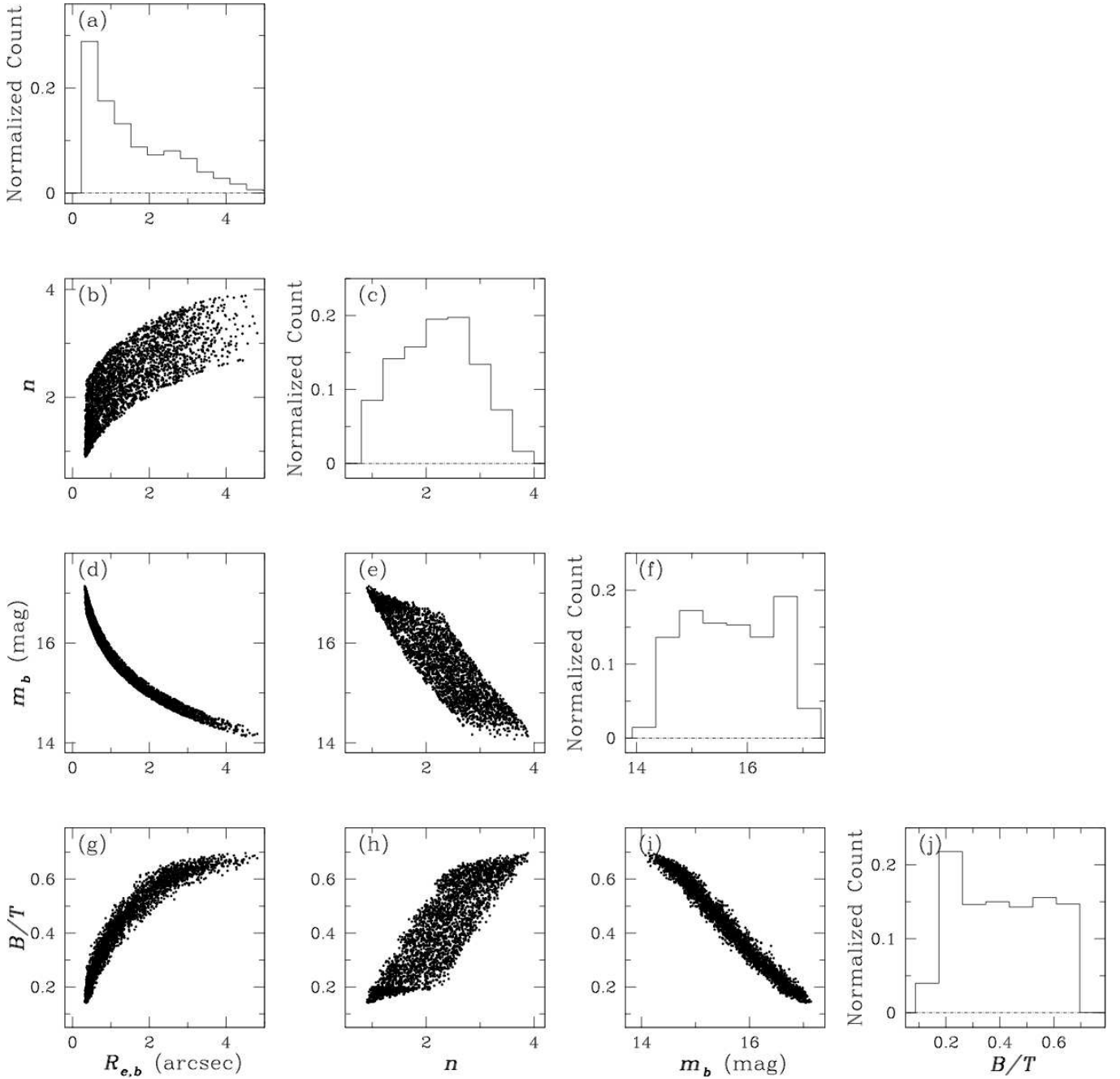}
  \caption{Properties of more than 2,000 simulated bulges rescaled to
    \emph{z} = 0.02 ($\sim$100 Mpc).
    Simulated galaxies have uniform distribution of
    $B/T$ in four bins ($<$0.2, 0.2--0.4, 0.4--0.6, and $>$0.6). 
   The correlation between magnitude ($m_B$) and effective radius
   ($R_e$) shown in panel (d) indicates that 
   bulges follow the Kormendy relation (Equation \ref{eq:KR}). Smaller
   bulges are less luminous, and therefore galaxies with the smallest
   $B/T$ have the smallest bulges (panel g). Panel (e) shows that  the S\'{e}rsic indices
   of bulges correlate with their total luminosity (Equation \ref{eq:M_n}). \label{fig:bulge}}
\end{figure*}

Our simulated images of fiducial local ($z\approx0$) galaxies will be
designed to mimic Sloan Digital Sky Survey (SDSS; York et al. 2000;
Stoughton et al. 2002) images taken in the $g$ band. 
Using the $V$-band Kormendy relation of \citep{Hamabe87} and assuming 
$g - V \approx 0.5$ for E/S0 galaxies (\citealt{Fukugita95}), our simulated bulges follow

\begin{equation}
\mu_e = 3.0 \ \log(R_e) + 20,
\label{eq:KR}
\end{equation}

\noindent where $R_e$ is expressed in Kpc and $\mu_e$
in mag ${\rm arcsec}^{-2}$ (Fig. \ref{fig:bulge}d). 

The simulated disk component follows the exponential light profile
(\citealt{Freeman70}). On the other hand, the S\'{e}rsic indices
(\citealt{Sersicc68}) of bulges correlate with their total luminosity
(\citealt{Fisher08}; \citealt{Graham08};
\citealt{Laurikainen10}). Using Equation 17 of \citet{Graham08}, which is given in the $B$ band, we assume $g-B = -0.7$ (appropriate for Sab galaxies; 
Fukugita et al. 1995) to obtain

\begin{figure*}[t]
  \centering
\includegraphics[width=135mm]{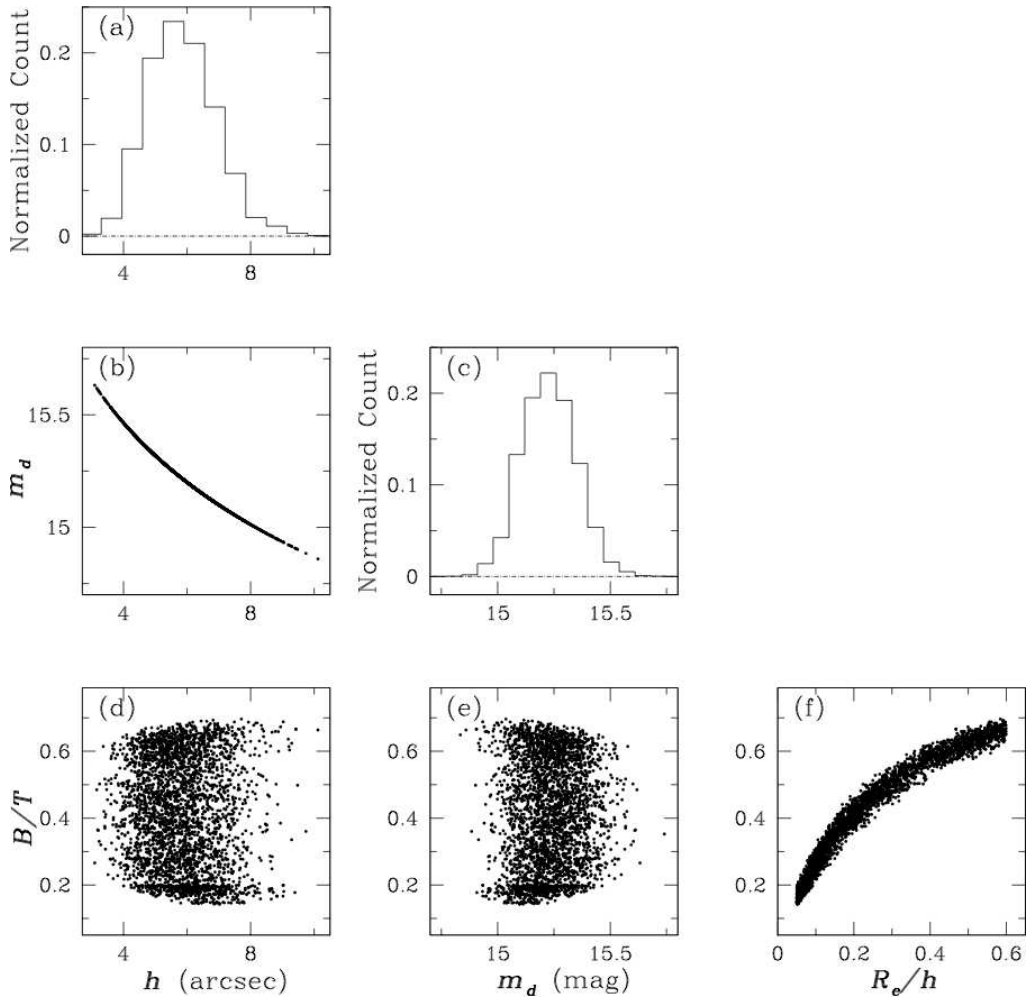}
  \caption{Properties of more than 2,000 simulated disks rescaled
    to \emph{z} = 0.02 ($\sim$100 Mpc).
    Simulated galaxies have uniform distribution of
    $B/T$ in four bins ($<$0.2, 0.2--0.4, 0.4--0.6, and $>$0.6). 
   The correlation between magnitude ($m_d$) and the disk scale length
   ($h$) demonstrates the scaling relation between the central surface brightness of the
   disk and its scale length. Panel (f) shows that galaxies with
lowest $B/T$ have the smallest $R_e$/$h$, in agreement with
previous studies. \label{fig:disk}}
\end{figure*}

\begin{figure*}[t]
  \centering
\includegraphics[width=170mm]{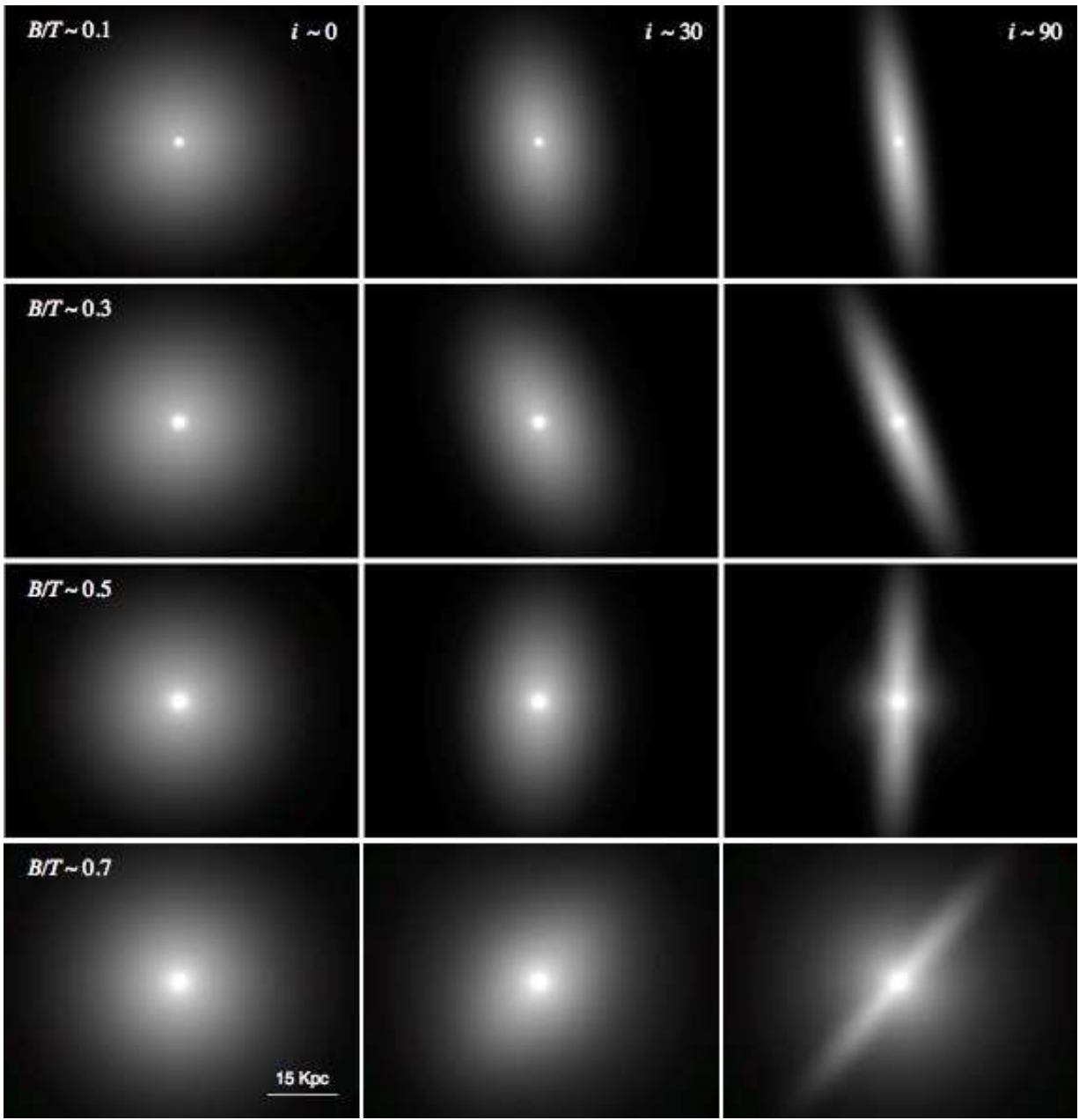}
  \caption{Simulated disk galaxies at four different values of $B/T$ and three
    inclination angles $i$. 
    No noise has been added to these
    images. The solid line in the lower left panel shows a scale of
    15 Kpc. The simulated images mimic SDSS data.\label{fig:example12}}
\end{figure*}

\begin{figure*}[t]
  \centering
\includegraphics[width=180mm]{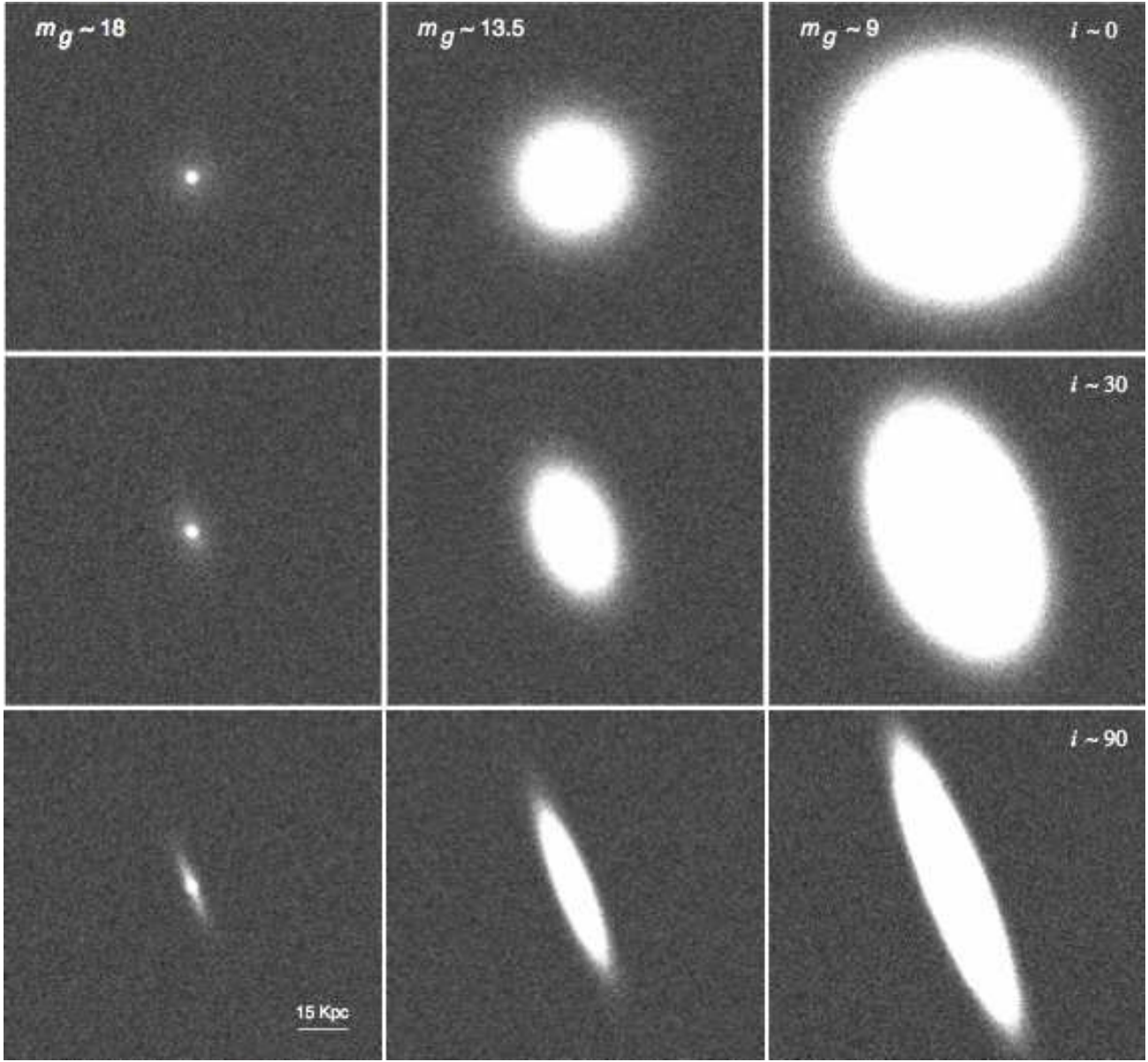}
  \caption{Simulated disk galaxies with $B/T$ $\approx$
0.3 at three different values of $S/N$, which correspond to $m_g$
$\approx$ 18, 13.5, and 9 mag, and three inclination
angles $i$. 
Galaxies with low $S/N$ look compact because the 
outer part of the disk is undetected.
The solid line in the lower left panel shows a scale of 15 Kpc. The
simulated images mimic SDSS data.\label{fig:example9}}
\end{figure*}

\begin{figure*}[t]
  \centering
\includegraphics[width=170mm]{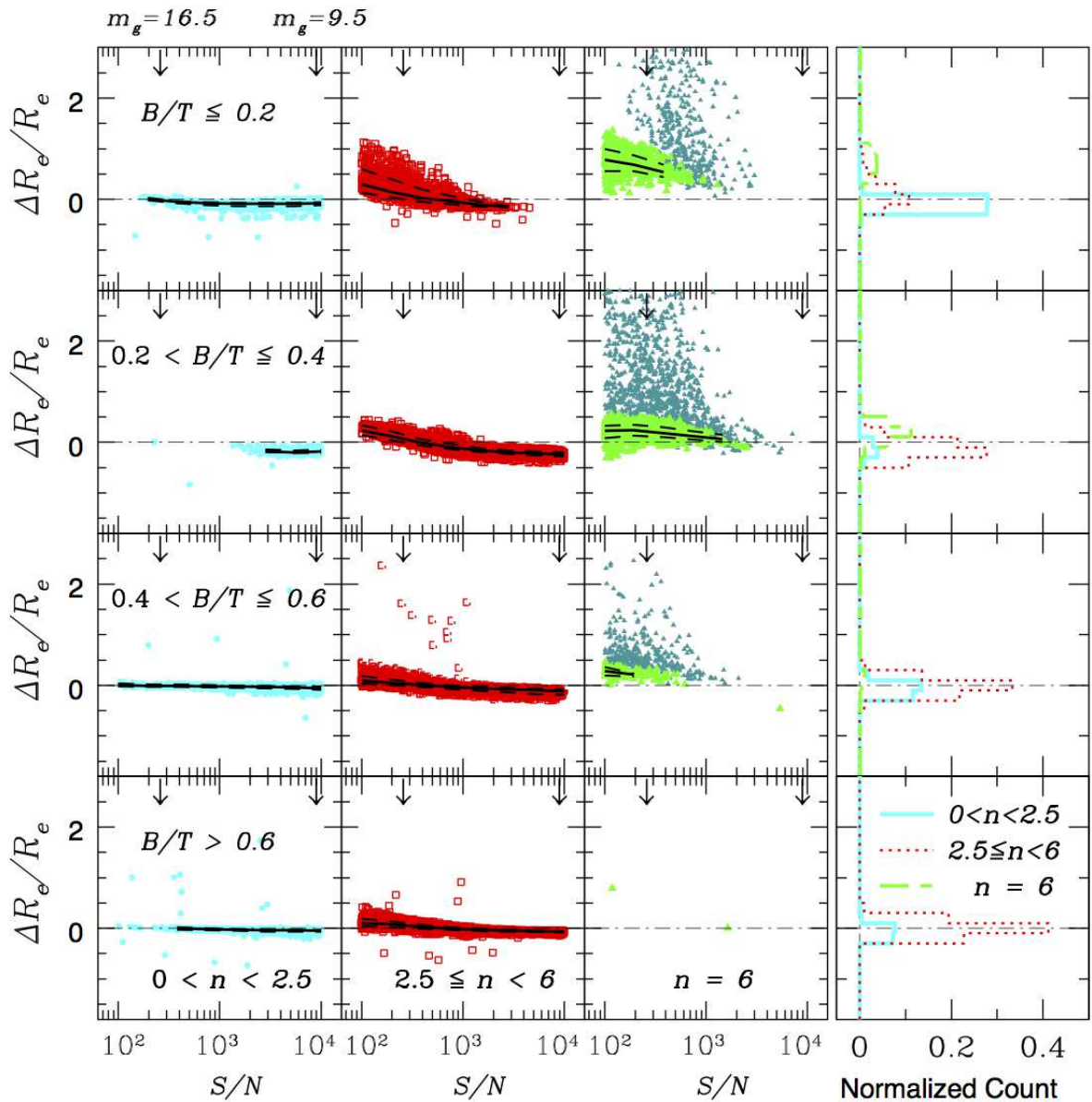}
  \caption{Fits of more than 30,000 \emph{local} disk galaxies with a single S\'{e}rsic
    component plus a sky component, which is left as a free parameter during the fit.
The sky component fits the background with a plane with a constant
slope and therefore can correct for any non-flatness, to first
order. Different rows show the offsets between the 
measured and the actual effective radius $R_e$ of galaxies with
different $B/T$; the different columns show three intervals in S\'{e}rsic index. 
Cyan solid circles/solid
lines, red open squares/dotted lines, and light green
solid triangles/dashed lines show the results of best-fit models
with $n$ $<$ 2.5, 2.5 $\leq$ $n$ $<$ 6, and
$n$ = 6, respectively. The dark green points show the models with $n$ $>$ 6.
The black solid and dashed lines indicate the median and 1$\sigma$ uncertainties of different
measurements. The downward-pointing arrows on the top of each subpanel
indicate the $S/N$ of a galaxy with $R_e$ $\approx$ 15 pixels,
$e$ = 0, and $m_g$ = 9.5 or 16.5. Comparing dark and light green points
shows that refitting single S\'{e}rsic fits with $n$ $>$ 6 by
fixing the S\'{e}rsic index to $n$ = 6 leads to significant improvement.\label{fig:1comp_Z0}}
\end{figure*}

\begin{figure*}[t]
 \centering
\includegraphics[width=135mm]{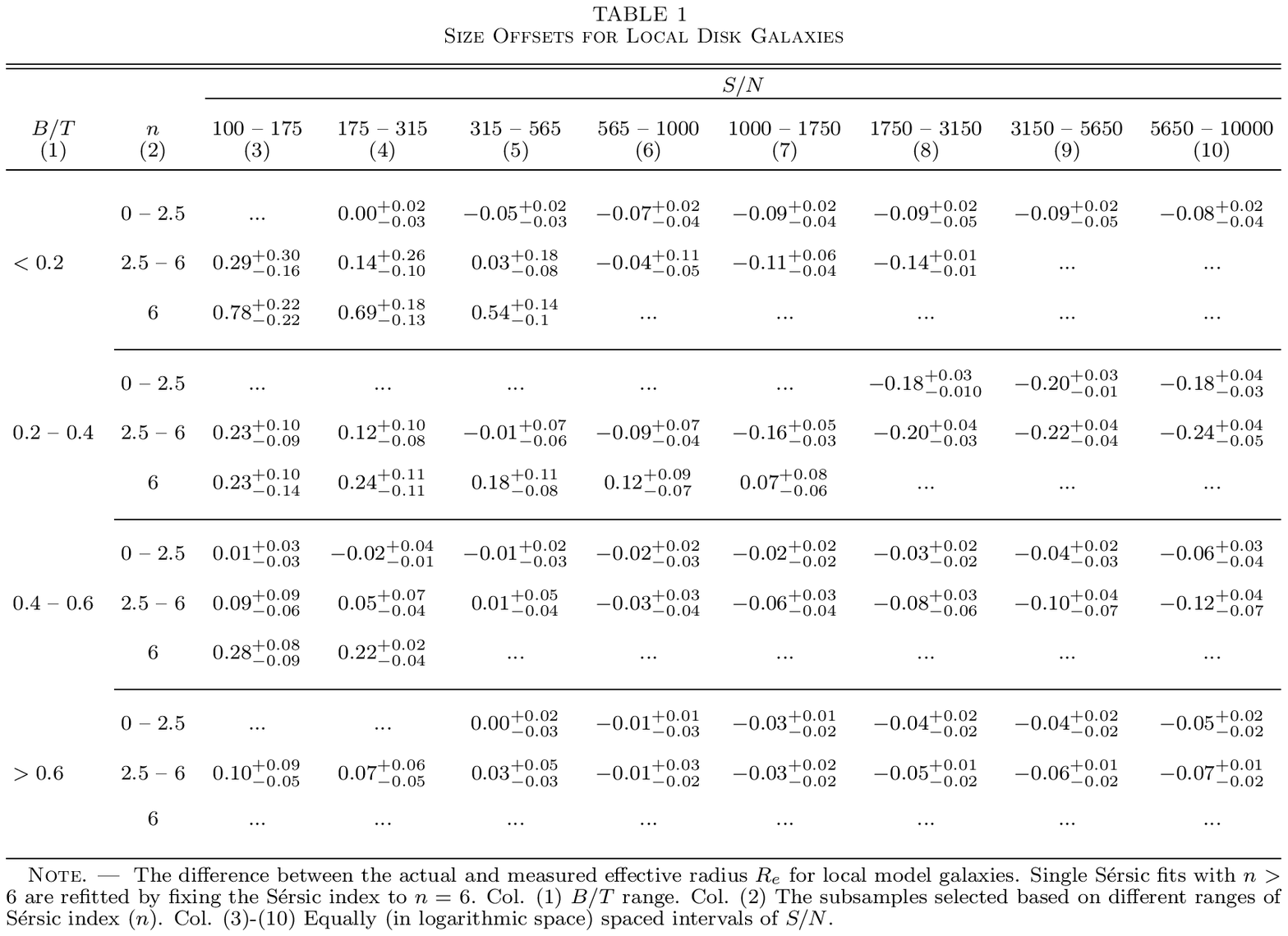}
\end{figure*}

\begin{figure*}[t]
  \centering
\includegraphics[width=135mm]{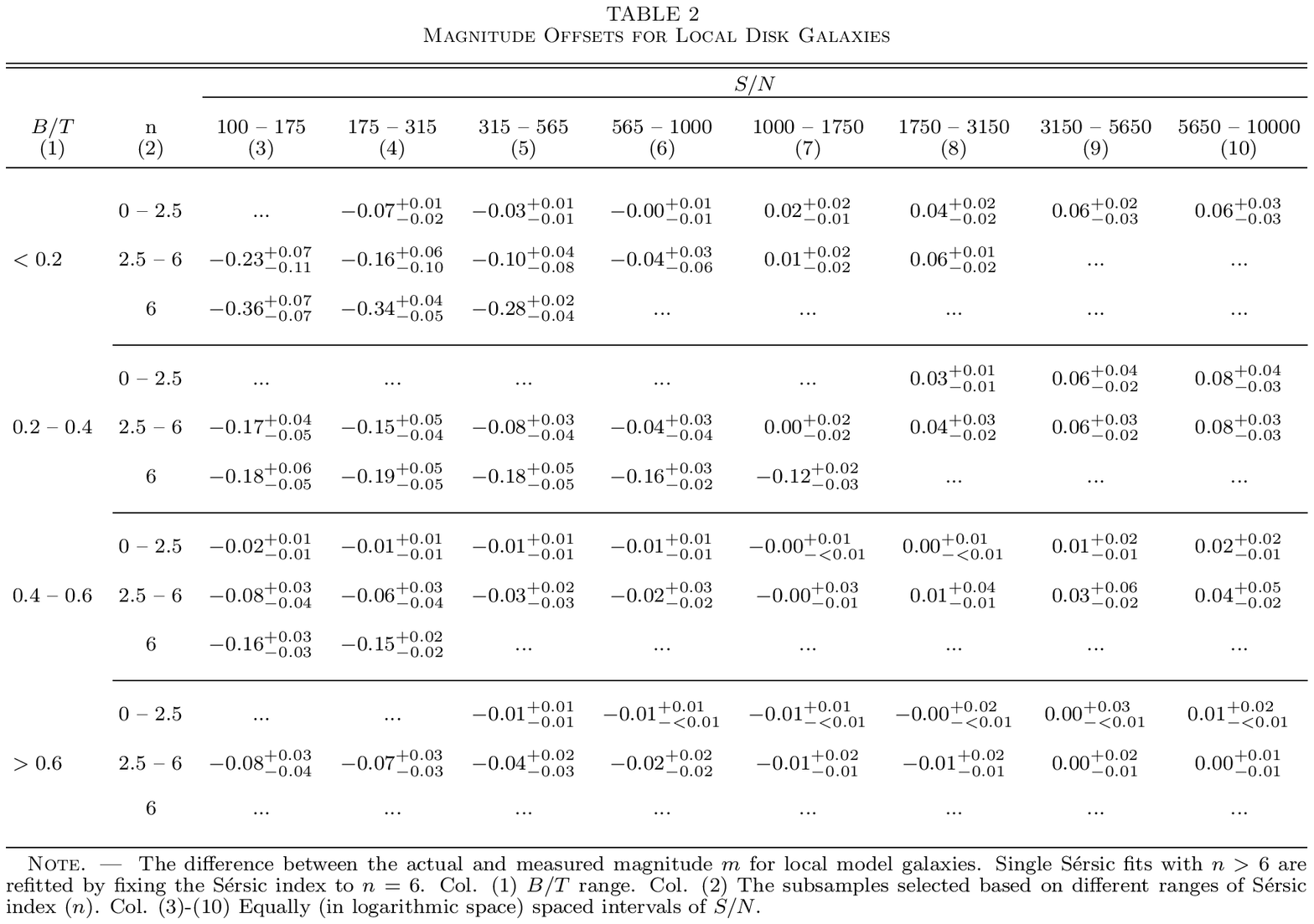}
\end{figure*}

\begin{figure*}[t]
  \centering
\includegraphics[width=170mm,angle=90]{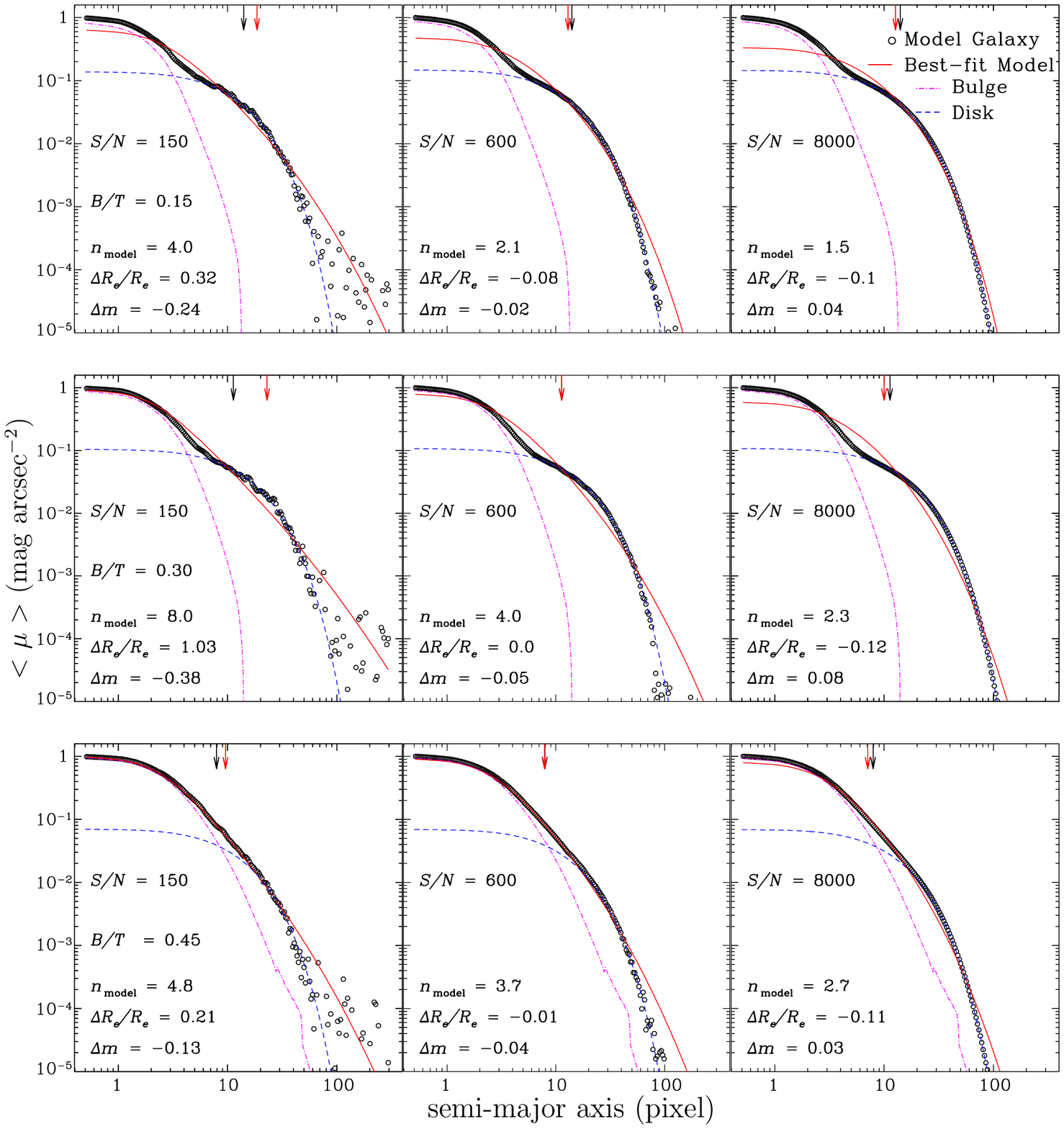}
  \caption{The effects of $S/N$
on the reliability of galaxy fitting. The different panels show
surface brightness profiles of simulated local galaxies with $B/T$ =
0.15, 0.3, and 0.45 at three different values of $S/N$.  Open black
and open red points show the light distributions of model galaxy and best-fit model
at any given $S/N$. Magenta dash-dotted and blue dashed lines show the
actual bulge and disk light distribution of the model
galaxy. The black and red downward arrows show the
  actual and single S\'{e}rsic effective radii, respectively. For model galaxies with $B/T$ $<$ 0.4, the large fraction of 
disk flux leads to systematic profile deviations in the outskirts; the
middle panels show that this effect is stronger for galaxies with 0.2
\lax\ $B/T$ \lax\ 0.4.  Lower $S/N$ leads to a situation where random
noise dominates over systematic profile deviations in the outskirts.
Our simulations show that the presence of a small bulge (lower $B/T$)
leads to relatively larger biases in size and luminosity using a
single-component fit. 
For model galaxies with $B/T$ $\geq$ 0.4, the dominance of the bulge reduces the
contribution of the disk to the total light distribution. The bottom
left panel shows that even at low
$S/N$, {\tt GALFIT}  can measure the size and 
total luminosity of the model galaxy reliably.\label{fig:1D_Z0}}
\end{figure*}

\begin{figure*}[t]
  \centering
\includegraphics[width=170mm]{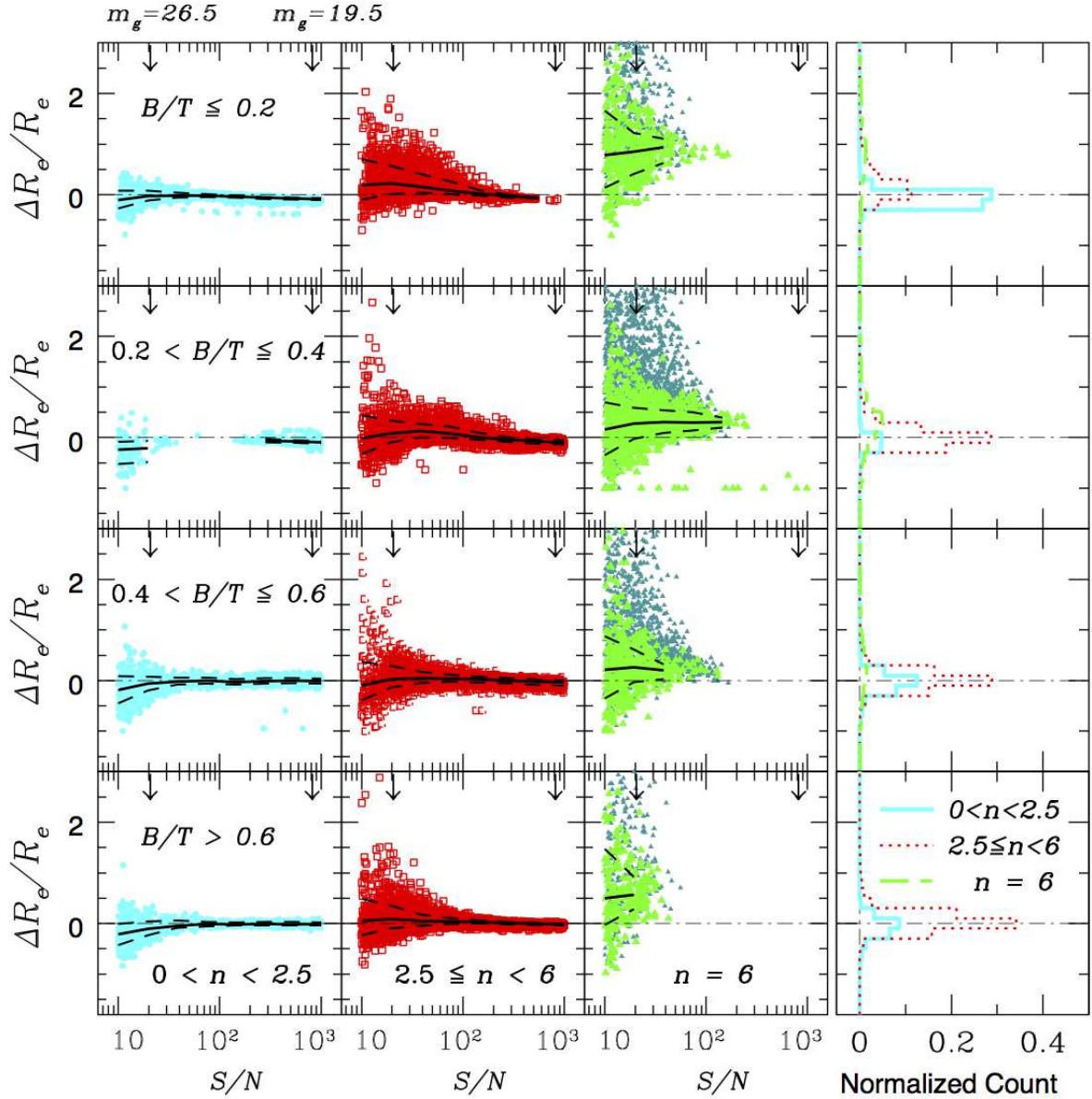}
  \caption{Single-component S\'{e}rsic fits of more than 30,000 disk
    galaxies \emph{rescaled} to \emph{z} = 2. All fits include a sky
    component, which is left as a free parameter. 
Different rows show the offsets between the
    measured and the actual effective radius $R_e$ of galaxies with different
    $B/T$; the different columns show three intervals in S\'{e}rsic index.
Cyan solid circles/solid
lines, red open squares/dotted lines, and light green
solid triangles/dashed lines show the results of best-fit models
with $n$ $<$ 2.5, 2.5 $\leq$ $n$ $<$ 6, and
$n$ = 6, respectively. The dark green points show the models with $n$ $>$ 6.
The black solid and dashed lines indicate the median and 1$\sigma$ uncertainties of different
measurements. The downward-pointing arrows on the top of each subpanel
indicate the $S/N$ of a galaxy with $R_e$ $\approx$ 5 pixels,
$e$ = 0, and $m_g$ = 19.5 or 26.5. Comparing dark and light green points
shows that refitting single S\'{e}rsic fits with $n$ $>$ 6 by
fixing the S\'{e}rsic index to $n$ = 6 leads to significant improvement.\label{fig:1comp_Z2}}
\end{figure*}

\begin{figure*}[t]
  \centering
\includegraphics[width=135mm]{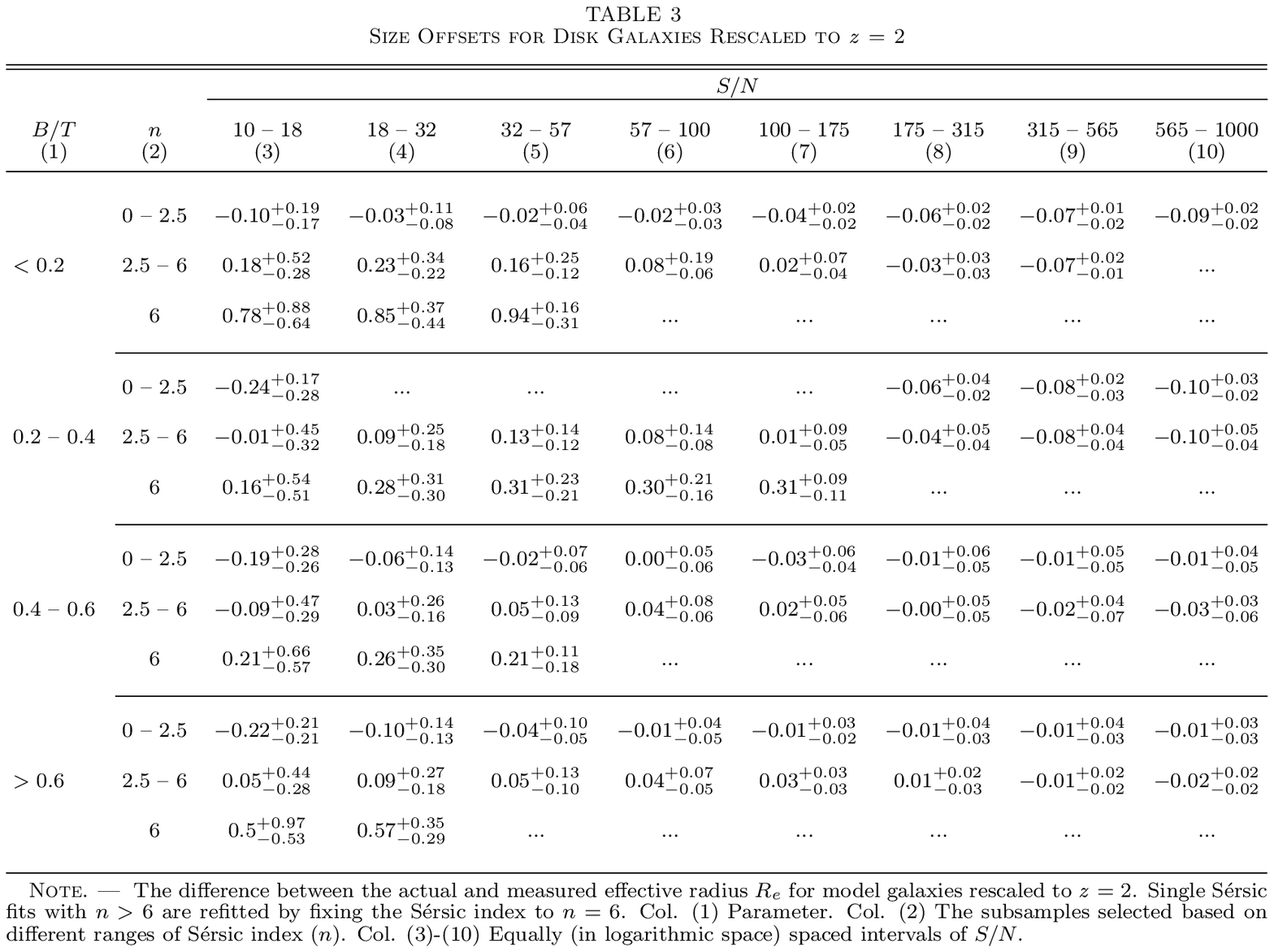}
\end{figure*}

\begin{figure*}[t]
  \centering
\includegraphics[width=135mm]{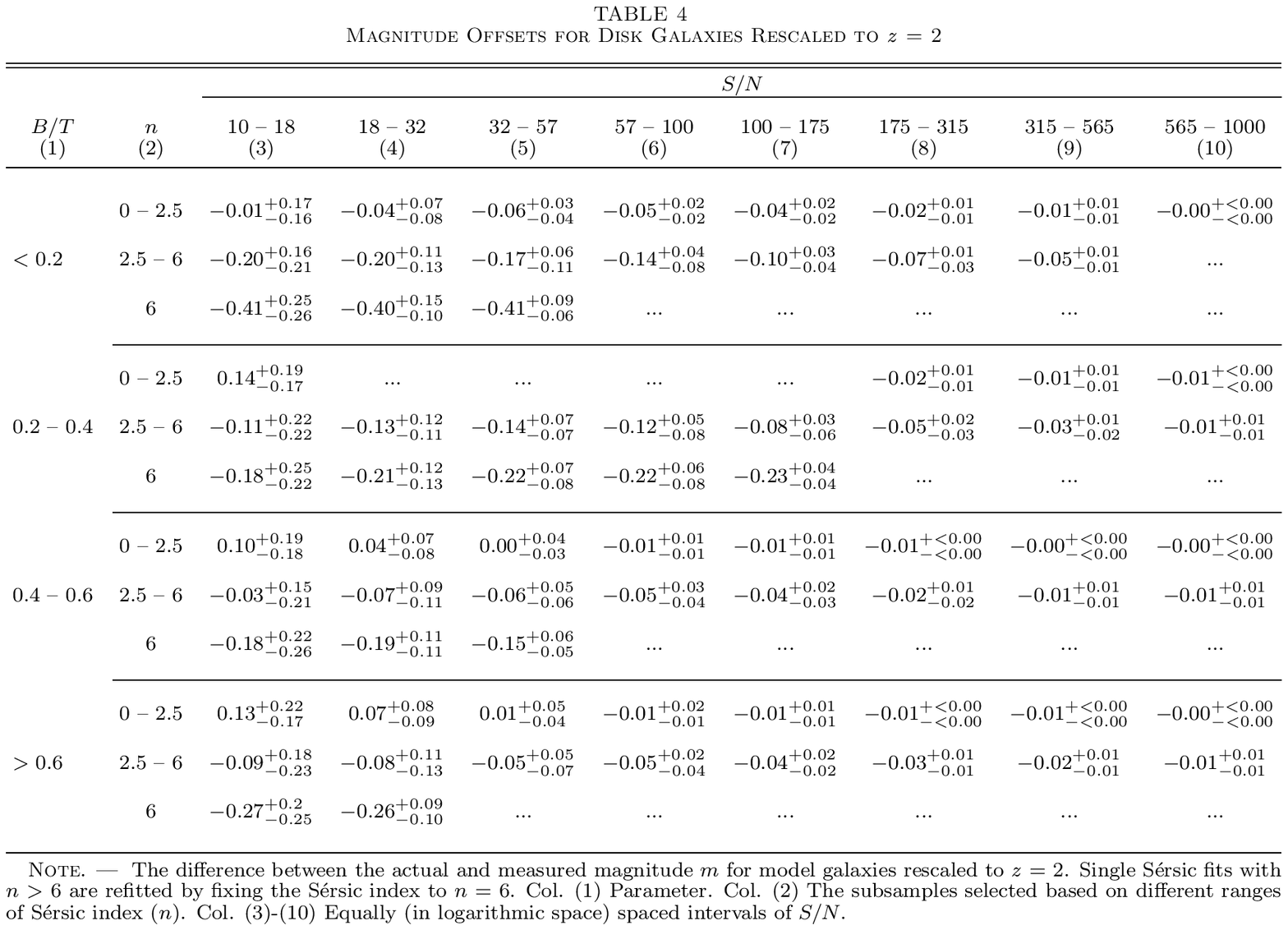}
\end{figure*}

\begin{figure*}[t]
  \centering
\includegraphics[width=120mm,angle=90]{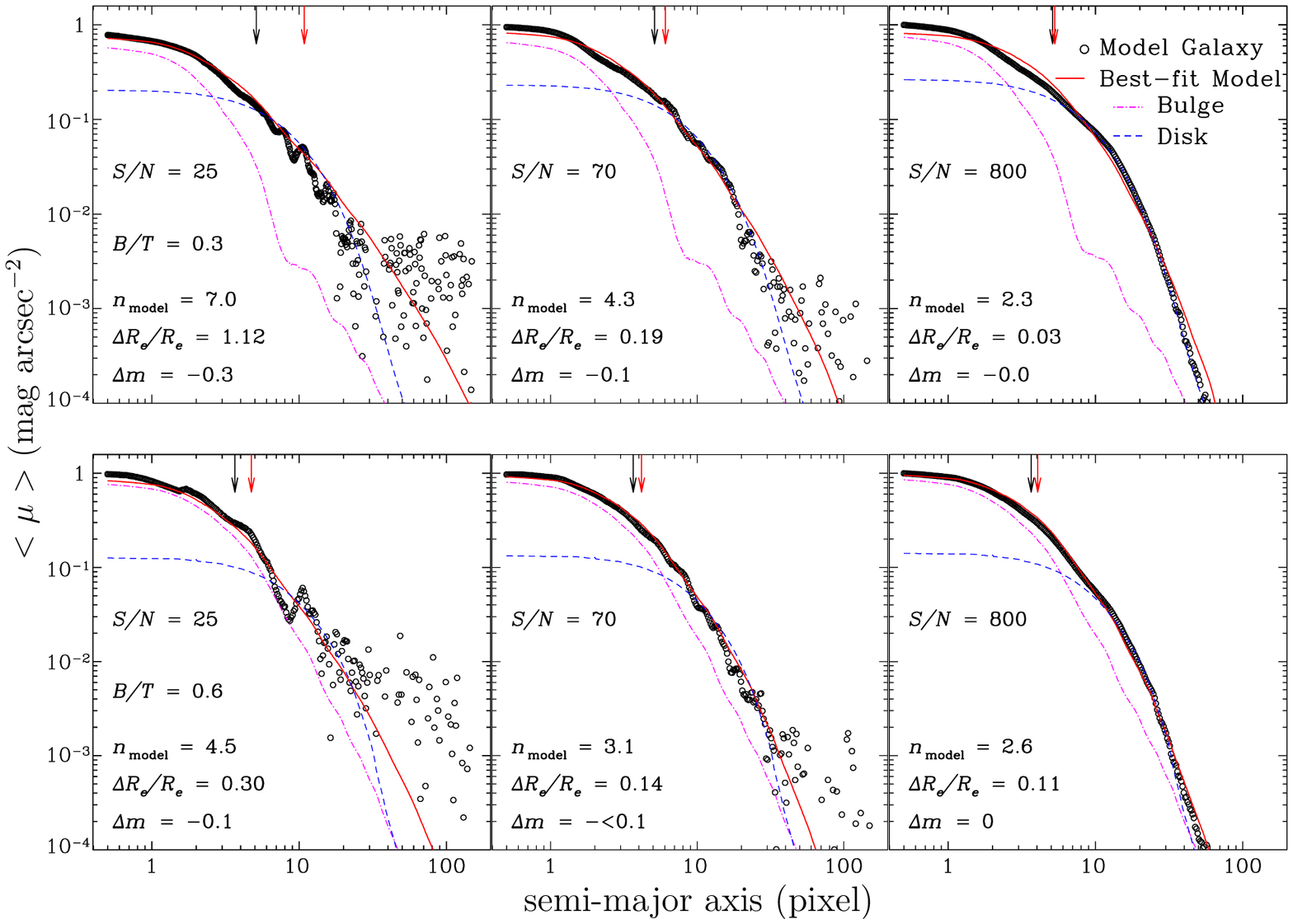}
  \caption{The effects of $S/N$ on the reliability of galaxy fitting.
    The different panels show surface brightness profiles of simulated
    galaxies  rescaled to \emph{z} = 2, with $B/T$ = 0.30 and 0.60 at
    three different values of $S/N$.  Open black and open red points
    show the light distributions of model galaxy and best-fit model at
    any given $S/N$. Magenta dash-dotted and blue dashed lines show
    the actual bulge and disk light distribution of the model galaxy. The black and red downward arrows show the
  actual and single S\'{e}rsic effective radii, respectively.  
Comparing this figure with Figure \ref{fig:1D_Z0} shows that rescaling
the local galaxies to higher redshift weakens the structural
non-homologies, and single-component models return more reliable fits
(top panels). Furthermore, for rescaled galaxies the $S/N$ can be so
low that it even affects the innermost part (i.e. bulge) of the galaxy
and therefore leads to unreliable single-component fits (lower panels). \label{fig:1D_Z2}} 
\end{figure*}

\begin{figure*}[t]
\centering
\includegraphics[width=180mm]{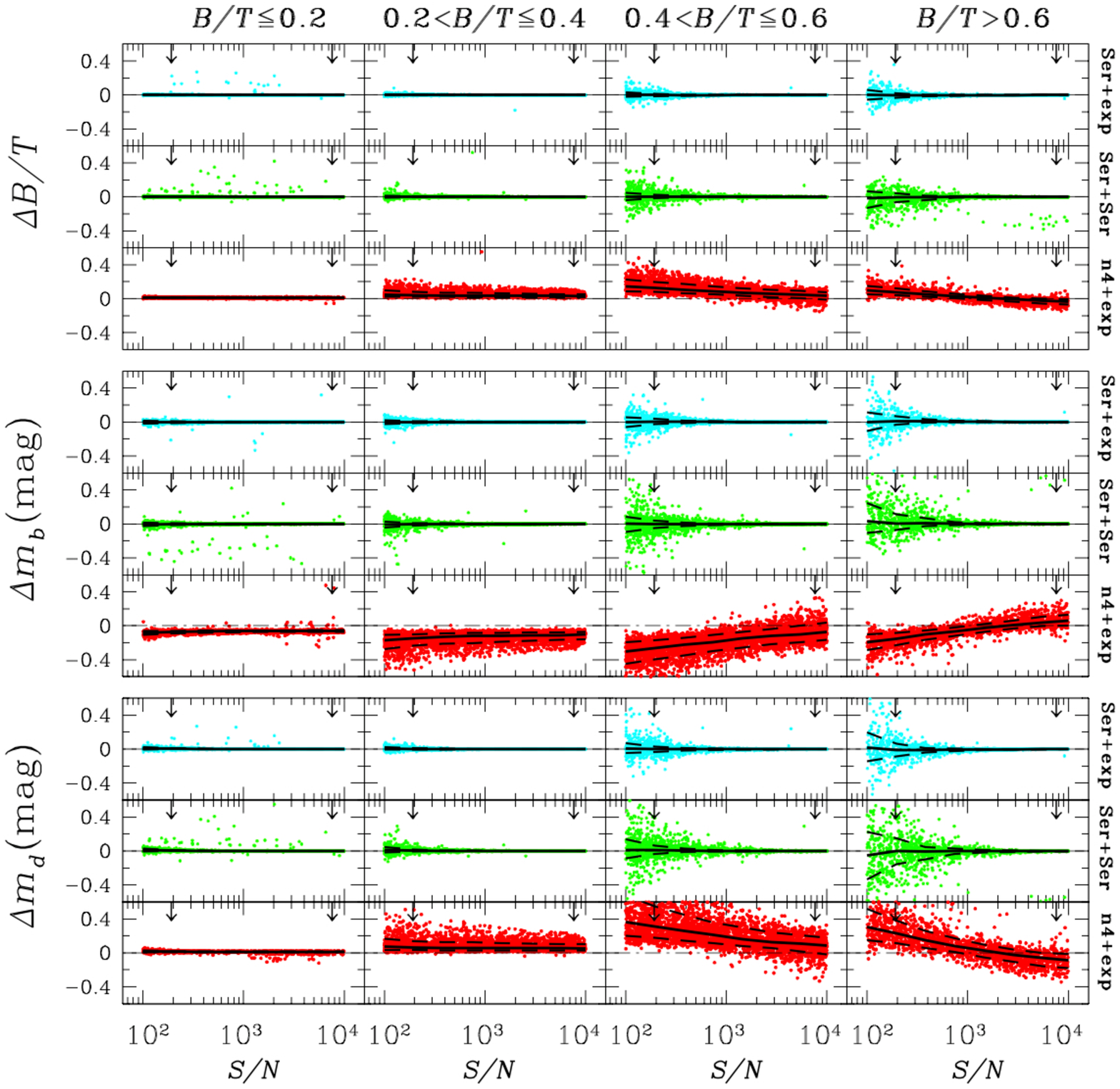}
\caption{Measuring total luminosity of the bulge and disk and $B/T$ of
  about 30,000 \emph{local} disk galaxies with three different models:
bulge with free  S\'{e}rsic + exponential disk (Ser+exp; cyan points), two S\'{e}rsic
  components (Ser+Ser; green points), and bulge with $n$4 + exponential disk
  ($n$4+exp; red points).  All fits include a sky
    component, which is left as a free parameter. Vertical panels show the results for
  different ranges of $B/T$, increasing from from left to
  right. Horizontal panels, from top to bottom, show the measurement
  offsets for $B/T$, bulge magnitude $m_b$, and disk magnitude $m_d$,
  respectively. The black solid and dashed lines indicate the median
  and 1$\sigma$ uncertainties of different measurements. The
  downward-pointing arrows on the top of each subpanel indicate the
  $S/N$ of a galaxy with $R_e$ $\approx$ 15 pixels, $e$ = 0, and $m_g$
  = 9.5 or 16.5. \label{fig:BTZ0}} 
\end{figure*}

\begin{figure*}[t]
\centering
\includegraphics[width=180mm]{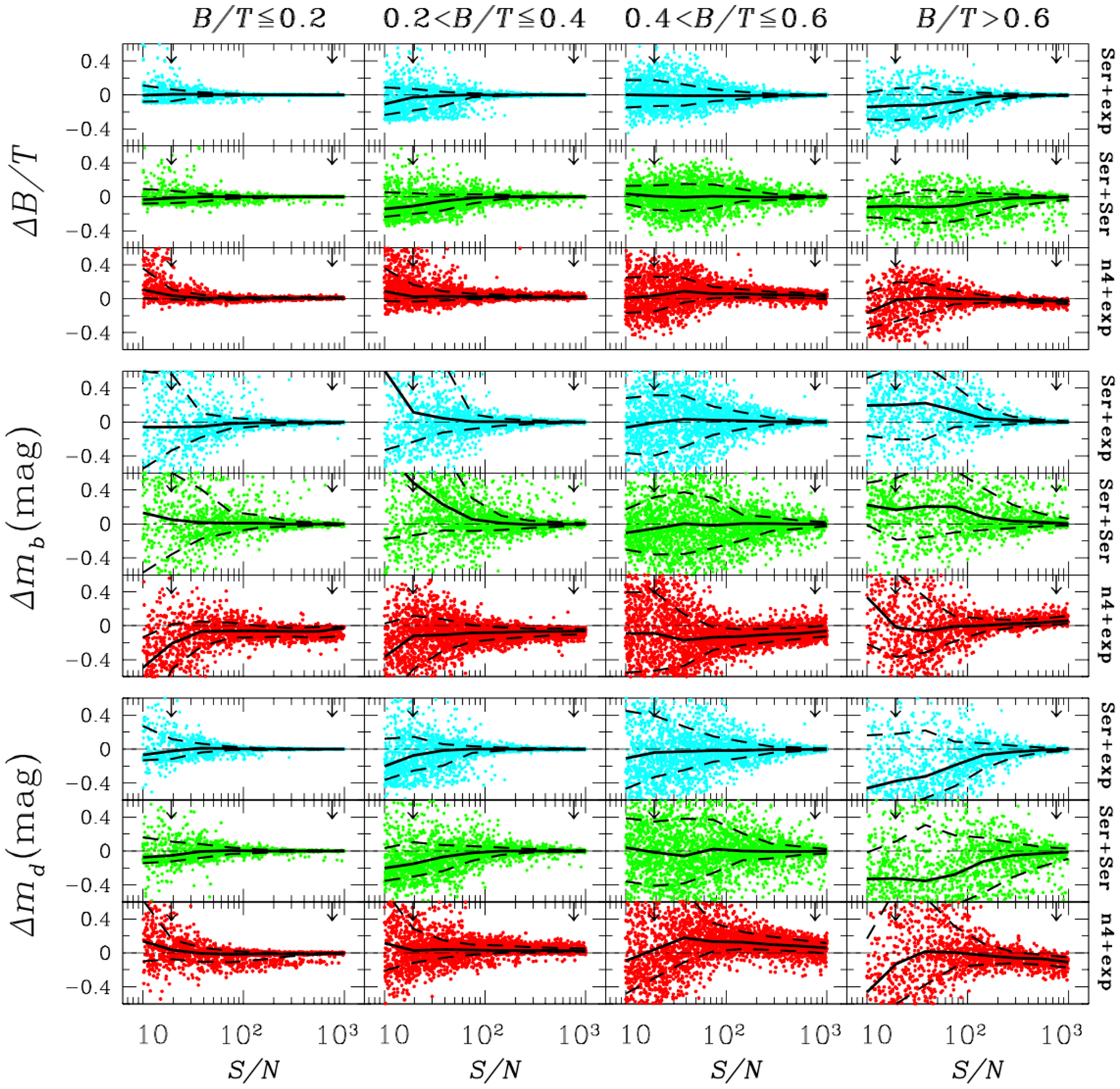}
\caption{Measuring total luminosity of the bulge and disk and $B/T$ of
  about 30,000 disk galaxies \emph{rescaled} to \emph{z} = 2 with
  three different models: bulge with free  S\'{e}rsic + exponential disk (Ser+exp; cyan points), two S\'{e}rsic
  components (Ser+Ser; green points), and bulge with $n$4 + exponential disk
  ($n$4+exp; red points). All fits include a sky
    component, which is left as a free parameter. Vertical panels show the
  results for different ranges of $B/T$, increasing from from left to
  right. Horizontal panels, from top to bottom, show the measurement
  offsets for $B/T$, bulge magnitude $m_b$, and disk magnitude $m_d$,
  respectively.  The black solid and dashed lines indicate the median
  and 1$\sigma$ uncertainties of different measurements. The
  downward-pointing arrows on the top of each subpanel indicate the
  $S/N$ of a galaxy with $R_e$ $\approx$ 5 pixels, $e$ = 0, and $m_H$
  = 19.5 or 26.5.\label{fig:BTZ2}} 
\end{figure*}

\begin{figure*}[t]
\centering
\includegraphics[width=180mm]{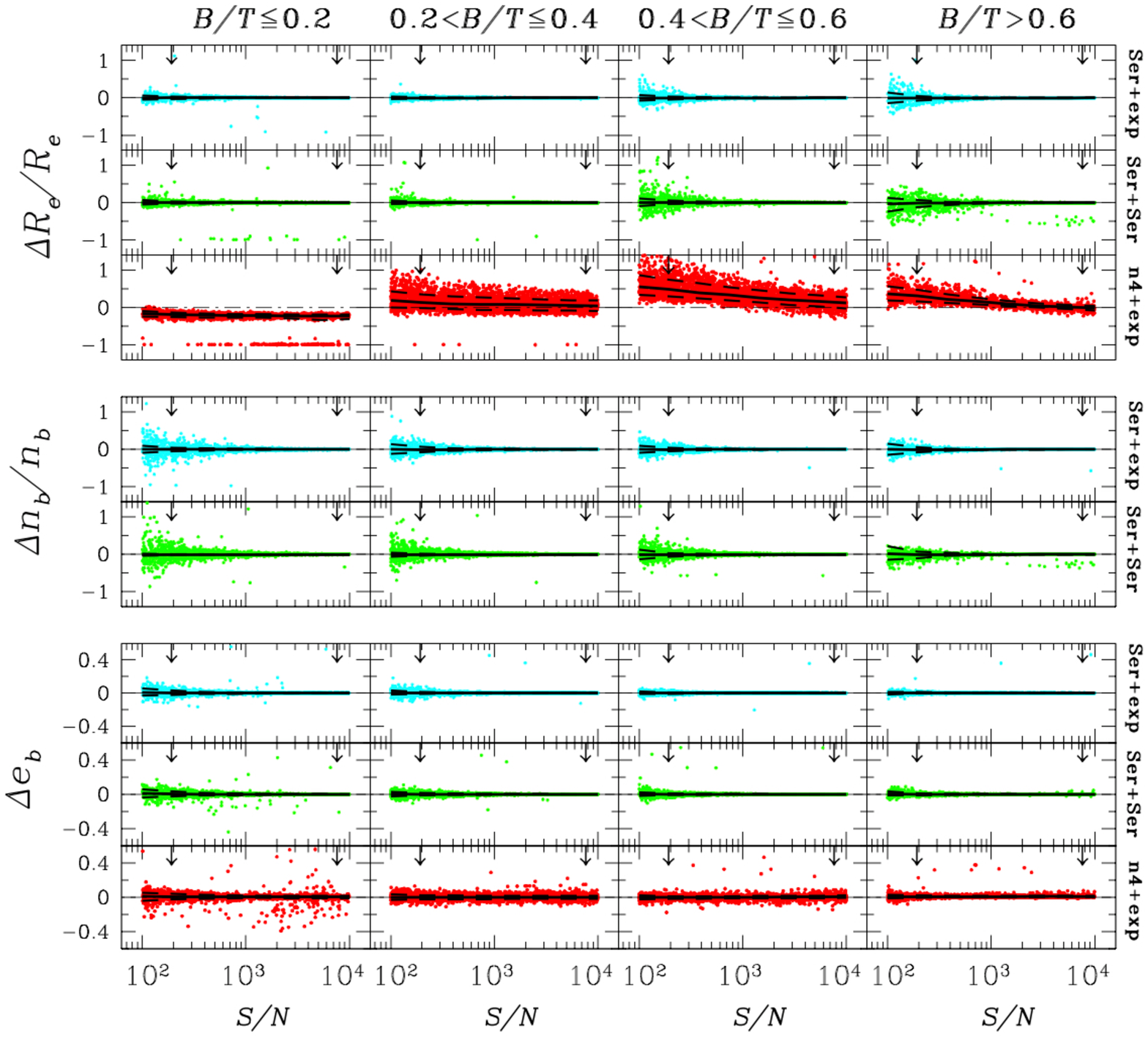}
\caption{Measuring bulge properties of about 30,000 \emph{local} disk
  galaxies with three different models:
bulge with free  S\'{e}rsic + exponential disk (Ser+exp; cyan points), two S\'{e}rsic
  components (Ser+Ser; green points), and bulge with $n$4 + exponential disk
  ($n$4+exp; red points). All fits include a sky
    component, which is left as a free parameter. 
Vertical panels show the results for different ranges of
  $B/T$, increasing from from left to right. Horizontal panels, from
  top to bottom, show the measurement offsets for bulge effective
  radius $R_e$, S\'ersic index $n_b$, and ellipticity $e_b$.  The
  black solid and dashed lines indicate the median and 1$\sigma$
  uncertainties of different measurements. The downward-pointing
  arrows on the top of each subpanel indicate the $S/N$ of a galaxy
  with $R_e$ $\approx$ 15 pixels, $e$ = 0, and $m_g$ = 9.5 or
  16.5.\label{fig:bulgeZ0}} 
\end{figure*}

\begin{figure*}[t]
\centering
\includegraphics[width=180mm]{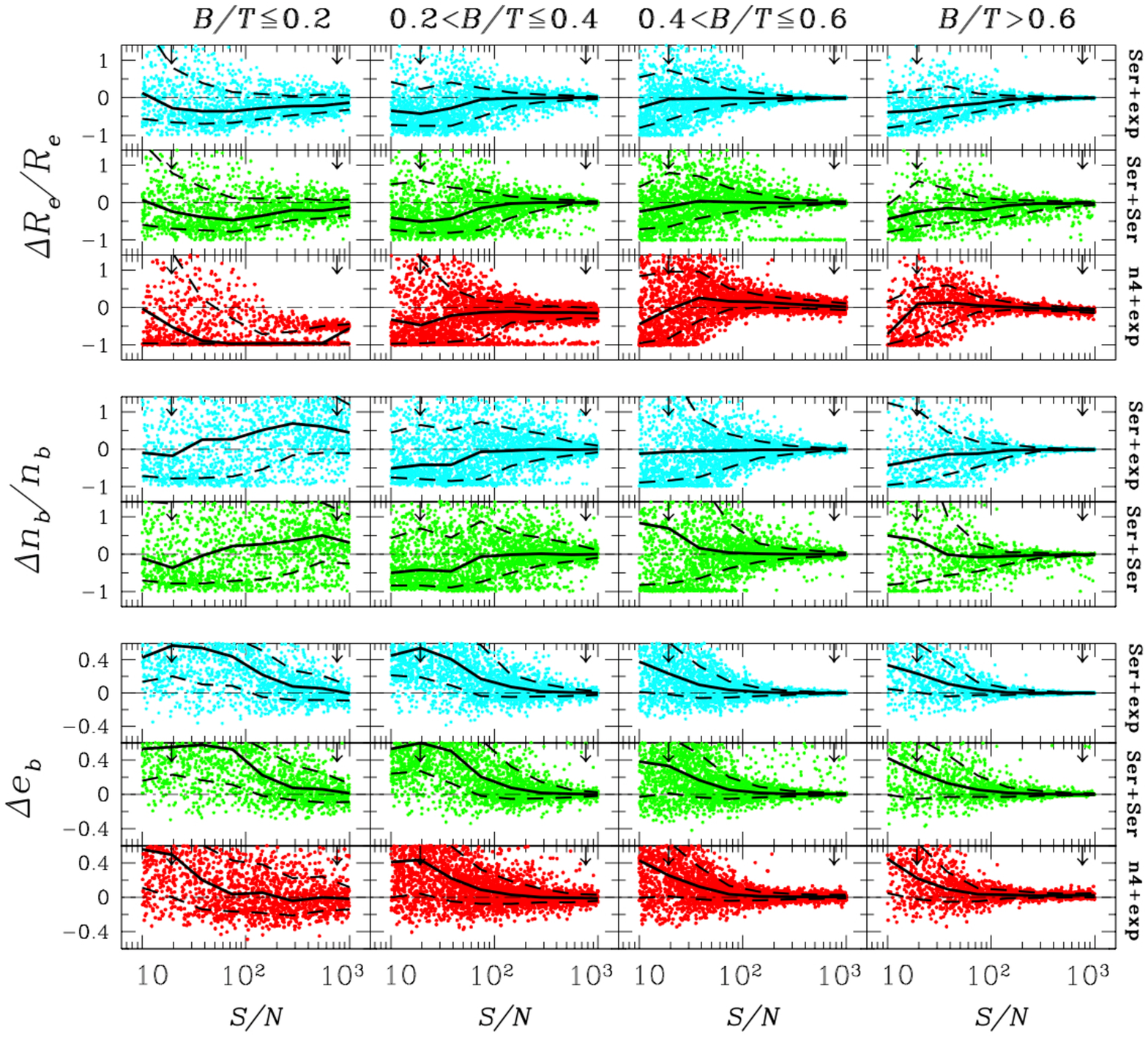}
\caption{Measuring bulge properties of about 30,000 disk galaxies
  \emph{rescaled} to \emph{z} = 2 with three different models:
bulge with free  S\'{e}rsic + exponential disk (Ser+exp; cyan points), two S\'{e}rsic
  components (Ser+Ser; green points), and bulge with $n$4 + exponential disk
  ($n$4+exp; red points). All fits include a sky
    component, which is left as a free parameter.  
Vertical panels show the results for
  different ranges of $B/T$, increasing from from left to
  right. Horizontal panels, from top to bottom, show the measurement
  offsets for bulge effective radius $R_e$, S\'ersic index $n_b$, and
  ellipticity $e_b$.  The black solid and dashed lines indicate the
  median and 1$\sigma$ uncertainties of different measurements. The
  downward-pointing arrows on the top of each subpanel indicate the
  $S/N$ of a galaxy with $R_e$ $\approx$ 5 pixels, $e$ = 0, and $m_H$
  = 19.5 or 26.5.\label{fig:bulgeZ2}} 
\end{figure*}

\begin{figure*}[t]
\centering
\includegraphics[width=180mm]{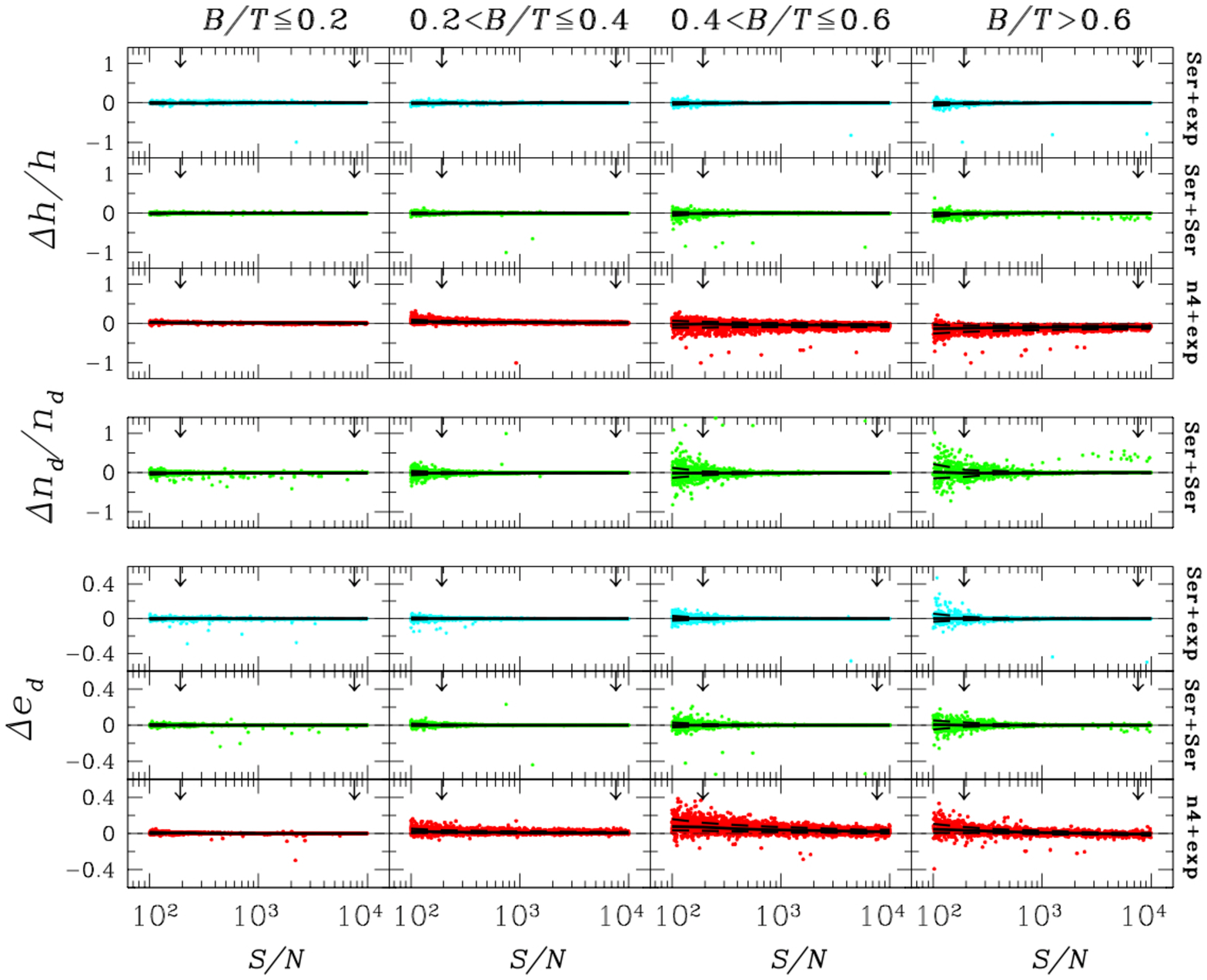}
\caption{Measuring disk properties of about 30,000 \emph{local} disk
  galaxies with three different models: bulge with free  S\'{e}rsic + exponential disk (Ser+exp; cyan points), two S\'{e}rsic
  components (Ser+Ser; green points), and bulge with $n$4 + exponential disk
  ($n$4+exp; red points). All fits include a sky
    component, which is left as a free parameter.  
Vertical panels show the results for different ranges of
  $B/T$, increasing from from left to right. Horizontal panels, from
  top to bottom, show the measurement offsets for 
disk scale length $h$, S\'ersic index $n_d$, and ellipticity $e_d$.
The black solid and dashed lines indicate the median and 1$\sigma$
uncertainties of different measurements. The downward-pointing arrows
on the top of each subpanel indicate the $S/N$ of a galaxy with $R_e$
$\approx$ 15 pixels, $e$ = 0, and $m_g$ = 9.5 or
16.5.\label{fig:diskZ0}} 
\end{figure*}

\begin{figure*}[t]
\centering
\includegraphics[width=180mm]{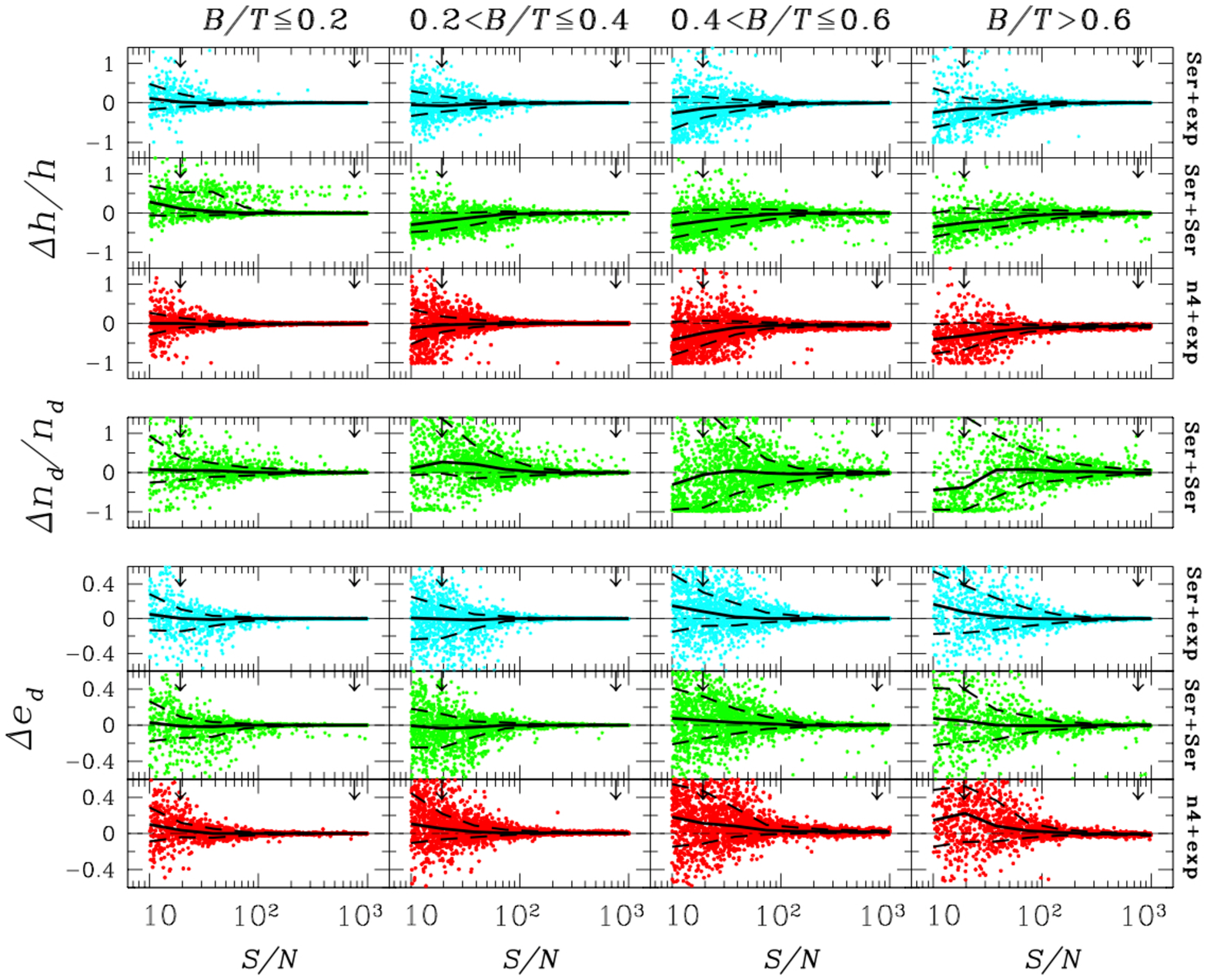}
\caption{Measuring disk properties of about 30,000 disk galaxies
  \emph{rescaled} to \emph{z} = 2 with three different models:
bulge with free  S\'{e}rsic + exponential disk (Ser+exp; cyan points), two S\'{e}rsic
  components (Ser+Ser; green points), and bulge with $n$4 + exponential disk
  ($n$4+exp; red points). All fits include a sky 
    component, which is left as a free parameter. Vertical panels show
  the results for different ranges of $B/T$, increasing from from left
  to right. Horizontal panels, from top to bottom, show the
  measurement offsets for disk scale length $h$, S\'ersic index $n_d$,
  and ellipticity $e_d$.  The black solid and dashed lines indicate the median and 1$\sigma$
uncertainties of different measurements. The downward-pointing arrows
on the top of each subpanel indicate the $S/N$ of a galaxy with $R_e$
$\approx$ 5 pixels, $e$ = 0, and $m_H$ = 19.5 or
26.5.\label{fig:diskZ2}} 
\end{figure*}

\begin{equation}
n = 10^{ -(15 + M_g) / 9.4},
\label{eq:M_n}
\end{equation}

\noindent where $M_g$ is the absolute $g$-band magnitude 
(Fig. \ref{fig:bulge}h). 

There are number of definitions of
  pseudo-bulges. Gadotti (2009) claims that ``pseudo-bulges can be
  distinguished from classical bulges as outliers in the Kormendy
  relation.'' Based on the fact that our simulated bulges follow
  the Kormendy relation, one might say that pseudo-bulges are not
  included in our sample. However, Fisher \& Drory (2008) argue that
  the main structural difference between pseudo and classical bulges
  is that the former has lower Sersic indices.  Without meaning to
  favor one definition over another, according to Fisher \& Drory
  (2008) definition, pseudo-bulges are present in our simulated
  sample. Definitions aside, as a prelude to the results, one can
  infer how well bulges can be fit (pseudo or not) regardless of the
  relation they follow because the accuracy of recovery can be
  summarized from direct measurables alone: luminosity, size, and $B/T$
  ratio. In short, if pseudo-bulges span a range that is lower in $S/N$,
  lower in angular resolution, and lower in $B/T$ than our simulations,
  then they would not be easily recovered at high-\emph{z},
  independent of which physical relations they follow.

The ellipticity, $e = 1 - b/a$, of bulges have a Gaussian
distribution that peaks around 0.2 (Fathi et al. 2003), whereas disks, as a result of different inclination angles, have $e < 0.8$ (\citealt{Ryden06}).
We assume disks and bulges have random orientations (position angles).

The simulated disks have a Gaussian distribution of central surface
brightness, with a mean and standard deviation of $\mu_0 = 21 \pm 0.3$
mag ${\rm arcsec}^{-2}$ (\citealt{Gadotti09}). 

These initial conditions lead
to a sample of model galaxies with 0.1 $<$ $B/T$ $<$ 0.7, with the majority
of the objects having 0.25 $<$ $B/T$ $<$ 0.5. There are more than 500
galaxies in each of the following bins of $B/T$: $<$0.2, 0.2--0.3,
0.4--0.3, 0.4--0.5, 0.5--0.6, and $>$0.6. This produces bulges with
$R_e \approx$ 0.4\asec --5\asec\ (1--12 pixels) and disks with $h
\approx$ 3.5\asec --10\asec (9--25 pixels). In this paper, the analyzed effective radii are circularized effective radii, $R_{e,{\rm circularized}} \equiv R_{e,{\rm GALFIT}} \times \sqrt{1-e}$; the same applies to the disk scale lengths.  Figures \ref{fig:bulge} and \ref{fig:disk} show the properties of the simulated bulges and disks. Figure \ref{fig:disk}f shows that galaxies with the smallest $B/T$ have the lowest $R_e/h$, in agreement with
previous studies (e.g., \citealt{MacArthur03}; \citealt{Fisher08}).

For this analysis, the sky backgrounds are simulated. Local and
high-\emph{z} simulated backgrounds have identical gain, pixel scale,
magnitude zero point, and average noise levels as in the SDSS and
CANDELS mosaic images, respectively. To obtain a robust determination of the sky background, the size of the simulated 
images should be at least 15 times larger than the effective radius of the galaxy 
(\citealt{Yoon11}; \citealt{Huang13a}). As the simulated images are meant to resemble 
SDSS data, the pixel scale is set to 0.396\asec, and the zero point of the images is set to 25 mag, 
typical for the SDSS $g$ band\footnote{http://www.sdss.org/dr7/algorithms/fluxcal.html}.
We convolve the simulated images with an empirical PSF constructed by stacking 30 bright stars selected from SDSS $g$-band images.

$S/N$ is a key factor for the robustness of morphological
analysis. Considering the fact that galaxies are extended objects, we
use the $S/N$ within the effective radius rather than the total $S/N$.  We define

\begin{equation}
S/N = \frac{f_{\rm galaxy}}{\sqrt{f_{\rm galaxy}+A\sigma^2}},
\label{eq:sn}
\end{equation}

\noindent where $A$ is the area within the effective radius and  $\sigma$ is the
sum of all possible sources of noise, including dark current, readout
noise, and shot noise from the sky. 

For each model galaxy, we generate 10 images with different noise levels over the
range 100 $< S/N <$ 10000; the $S/N$ values are uniformly distributed in logarithmic 
space. In total, we generate more than 5000 images in each $B/T$ bin, and more than 
30,000 over the whole range of $B/T$.  For the typical noise level and exposure time 
of SDSS images, the above range of $S/N$ values corresponds to simulated galaxies with 
$8.5 < m_g < 18.5$. 

Figure \ref{fig:example12} shows simulated disk galaxies with four different values of 
$B/T$ and three inclination angles.  Figure \ref{fig:example9} gives an example of a 
simulated galaxy with $B/T \approx 0.3$ at three different levels of $S/N$, which 
correspond to $m_g$ $\approx$ 18, 13.5, and 9 mag, and three inclination angles. 
Objects with low $S/N$ look like compact galaxies due to the missing light in the 
outer part of the disk. 

One of the main goals of this paper is to examine the reliability of model fits of 
high-redshift massive galaxies using one or two S\'{e}rsic components.
These high-redshift galaxies are generally smaller than local
galaxies, have been observed at lower 
$S/N$, and have been subject to surface brightness dimming.
We simply rescale the size and $S/N$ as if the local galaxies are
observed at \emph{z} $\approx$ 2 without taking the size evolution or
stellar population evolution into account. This leads to simulated
galaxies with bulges with 0.02\asec\ $< R_e <$ 
0.25\asec\ (0.3--4 pixels) and disks with
0.17\asec\ $< h <$ 0.5\asec\ (3--8 pixels). The rescaled model
galaxies are simulated at 10 $< S/N <$ 1000 (uniformly distributed
in logarithmic space), and their images mimic  
the properties of the CANDELS\footnote{\tt
  http://candels.ucolick.org/} (\citealt{Grogin11};
\citealt{Koekemoer11}) UDS (UKIDSS Ultra-Deep Survey;
\citealt{UKIDSS}) mosaic image. For images with the noise level of
those acquired using {\it HST}'s Wide Field Camera 3 (WFC3) in the
$H_{160}$ filter (\citealt{Koekemoer11}), this  range of $S/N$ for the
rescaled model galaxies is equivalent to galaxies with  
$18.5 < H_{160} < 28.5$ observed for one orbit with the {\it
  HST}. Note that our simulated magnitudes covers
  a range well below the detection limit of high-redshift studies in
  near-infrared (including CANDELS) and thus provides the chance to
  detect the $S/N$ levels at which reliable morphological measurements
  can be achieved. The simulated images are convolved with the hybrid $H_{160}$ PSF from
CANDELS UDS  (\citealt{vanderWel12}).

Once the galaxies are simulated, we fit each of them in two ways: (1)
single S\'{e}rsic component plus a sky component and (2) a S\'{e}rsic
component (to represent a bulge), an exponential component (to
represent a disk), and a sky component.   The sky component is fitted
as a plane with a constant slope, to correct for any residual
non-flatness, to first order, for all the following analysis. All parameters are left
  free (including the sky component parameters) except when the disk
  or/and bulge S\'{e}rsic index is fixed.  Besides, no constraint file
  (for limiting the parameter space) is used for {\tt GALFIT} fits .

\section{Results}

\subsection{Fitting Disk Galaxies with Single S\'ersic Component}

In order to quantify the reliability of our model fits, we
need to determine the intrinsic size of a simulated galaxy. Davari et
al. 2014 (their Fig. 3) show that this can be done robustly by using
{\tt IRAF}/{\tt ellipse} (\citealt{Jedrzejewski87}) and constructing
the galaxy curve-of-growth on the input, noiseless images.

\subsubsection{Local Galaxies}

We first discuss the results for single-component fits, the method
most widely adopted in the literature to obtain the total luminosity
and basic structural parameters of galaxies. Figure \ref{fig:1comp_Z0}
shows the difference between actual and measured effective radii as a
function of $S/N$, for more than 30,000 model galaxies. The horizontal
panels give the results at a given $B/T$, while the vertical panels
give the results for different values of the S\'{e}rsic index $n$.   

As shown by \citet{Davari14}, the fits with $n$ $>$ 6 are
generally unreliable. This comes from the fact that S\'{e}rsic
profiles with $n$ $>$ 4 have a long tail, which can be confused by the
background signal and its associated noise. This generally causes
overestimation of the total flux and size of
a galaxy. We refit these galaxies with $n$ $>$ 6
by fixing $n$ to 6, to minimize potential biases caused by these
unreliable fits. Comparing dark and light green points in Figure
\ref{fig:1comp_Z0} demonstrates the significant improvement
in the size determination for best-fit models with high S\'{e}rsic
indices. All the reported results allow the sky
component to be free. However, when $n$=6 is forced, the sky value naturally has 
to increase to compensate for the lower index value.  The increase in
the sky value, however, is not necessarily large.  For large
S\'{e}rsic indices ($n >$ 4) it is typical that a small change in sky
value can produce a large swing in the index value.  

We note that fixing $n$ to different values leads to
  different biases for galaxies at different 
$B/T$ range. For example, fixing $n$ to 2.5 will reduce biases for
galaxies with 0.2 $ < B/T <$ 0.4. However, when $n$ grows larger than 4, it is
common to hold it fixed to 4.  For our sample disky galaxies, holding
$n$ fixed to 4 should reduce the systematics compared to $n$=6, but
the differences are not large.  

\paragraph{$B/T$ Dependence of Fitted S\'{e}rsic Indices} 

The histograms in Figure \ref{fig:1comp_Z0} show that the fraction of
best-fit models with 2.5 $\leq$ $n$ $<$ 6 is higher for model galaxies
with higher $B/T$. This is expected, as previous studies have shown
that the light distributions of later type galaxies resemble
exponential disks ($n$ = 1), whereas bulge-dominated galaxies have
light distributions close to $n$ = 4
(\citealt{Graham08}).  On the other hand, higher
background noise level affects the model galaxies with smaller $B/T$
the most.  This is evidenced by the fraction of single-component fits
that yield high S\'{e}rsic indices ($n > 6$).  As mentioned above,
fits with $n > 6$ are generally unreliable and can lead to significant
systematic offsets. This, in part, comes from the fact that S\'{e}rsic
profiles with large $n$, which have a steep central part and a long
tail at large radii, are very sensitive to changes in the central and
outermost parts of the galaxy.  The fraction of galaxies with $n \geq
6$ are 15$\%$, 25$\%$, 2$\%$, and $<$1$\%$ for galaxies with $B/T
<0.2$, 0.2--0.4, 0.4--0.6, and $>0.6$, respectively.  

\paragraph{$S/N$ Dependence of Systematics}

The higher fraction of unreliable fits at lower $S/N$ for model galaxies with
$B/T < 0.4$ is due to their morphological properties.  This is
demonstrated in Figure \ref{fig:1D_Z0}, which illustrates the effects
of $S/N$ on the reliability of the fits.  Light distributions of model
galaxies with $B/T$ = 0.15, 0.30, and 0.45 are shown at three
different values of $S/N$.  
For model galaxies with $B/T$ $<$ 0.4 a considerable fraction of the
flux resides in the disk, which leads to systematic profile deviations
in the outskirts. Figure \ref{fig:1D_Z0} shows that this effect is stronger for galaxies with 0.2 \lax\ $B/T$ \lax\ 0.4.
Lower $S/N$ leads to a situation where random noise dominates over
systematic profile deviations in the outskirts.  As the contribution
of ${\chi}^2$ between the outer and inner region is asymmetric, it is
generally difficult to know a priori how the S\'{e}rsic index would
behave when fitting a single-component profile to a multi-component
model.  Our simulations show that the presence of a
small bulge (lower $B/T$) preferentially leads to relatively larger
biases in size and luminosity using a single-component fit.  Only in
the largest $B/T$ and the highest $S/N$ does the fit have a (slightly)
opposite effect.  In model galaxies with $B/T \geq 0.4$, the prominence of the bulge
reduces the contribution of the disk to the overall light
distribution. The bottom left panel shows that even at very low $S/N$
{\tt GALFIT}  can still reliably measure the size and total luminosity
of the model galaxy. This is in agreement with Davari et al. (2014),
who find that single-component fits of early-type galaxies can
accurately measure their structural properties and total luminosities
over a wide range of $S/N$. 

Note that although single-component fits to model galaxies with $0.2 <
B/T < 0.4$  have the lowest fraction of best-fit models with $n <
2.5$, at high $S/N$ their fraction is similar to that of model
galaxies with $0.4 < B/T < 0.6$ and higher than that for $B/T >
0.6$. Therefore at higher $S/N$, as expected, the fraction of best-fit
models with $n < 2.5$ is higher for model galaxies with lower $B/T$.

Tables 1 and 2 summarize the results of the single-component fits.  As
mentioned before, at low $S/N$, it is more likely to overestimate the
total luminosity and size. On the other hand, at high $S/N$ the
non-homology in the light distribution of these two-component galaxies
leads to underestimaton of the total luminosity.  The offsets are
larger for model galaxies with $0.2 < B/T < 0.4$, as these galaxies
have the most prominent no-homology in their surface brightness light
distribution (Fig. \ref{fig:1D_Z0}). And, as expected, the offsets are
smaller for bulge-dominated galaxies because of the negligible
contribution of the disk component to the overall light profle.  

For the range of $S/N$ in which typical local disk galaxies are
observed (i.e. a few thousands), systematic offsets are less than
10\%--20\% for size measurements and \lax 0.1 mag for luminosity
measurements. Thus, single-component  S\'{e}rsic fits using {\tt
  GALFIT} give reliable sizes and luminosities for typical local disk
galaxies. 

\paragraph{Sky Estimation Effect on Systematics}

Estimating the sky value is another key factor in morphological
analysis, especially for extended objects like galaxies (e.g.,
\citealt{MacArthur03}; \citealt{Erwin08}; \citealt{Bernardi10}; \citealt{Yoon11}).
Davari et al. (2014; their Fig. 9)  show that the sky value can be
measured with an accuracy better than 0.1$\%$ for early-type galaxies.
That finding holds for bulge-dominated disk, galaxies, but it may not be the case for
 more disk-dominated systems. As discussed above, low-$S/N$,
 disk-dominated galaxies paradoxically may be better fit by a model 
with a high S\'{e}rsic index, whose extended wings contribute little to ${\chi}^2$ compared to improvements in the fit toward the center.
 In order to examine the effect of sky determination, all model galaxies are
fit two ways: (1) by letting the sky be a free component in the fit and (2)
by fixing the sky value to the actual background value measured independently.  
We found that at high $S/N$ {\tt GALFIT} can measure the sky value accurately when 
enough background pixels are available. (Note that the simulated images are 
at least 15 times larger than the $R_e$ of the model galaxy.) But at low $S/N$ 
(less than a few hundreds) and for galaxies with low $B/T$, fixing the sky 
value decreases the systematic offsets by $\sim 10$\% for 
$2.5 \leq n < 6$ and by $\sim 20$\% for $n \geq 6$.

\subsubsection{ Galaxies Rescaled to \emph{z} = 2 }

Figure \ref{fig:1comp_Z2} shows the results of fitting disk galaxies
scaled to $z = 2$  with a single S\'{e}rsic component. The trends are similar to
the results for local disk galaxies (Fig. \ref{fig:1comp_Z0}), but
Tables 3 and 4 show that at high $S/N$ the systematic offsets are
smaller for rescaled galaxies. As a result of rescaling the size of
local galaxies to mimic galaxies at higher redshift, the
structural non-homologies are mostly washed out and single-component
fits, in fact, return more reliable fits. This can be seen in Figure
\ref{fig:1D_Z0} (middle panels) and Figure \ref{fig:1D_Z2} (top panels). 

Another noticeable difference between the results of local and
rescaled model galaxies is the higher fraction of single-component
fits with $n \geq 6$ for more bulge-dominated galaxies. We consider
these fits to be unreliable.  The fraction of unreliable fits are
8$\%$, 20$\%$, 7$\%$, and $<$4$\%$ for galaxies with $B/T <0.2$,
0.2--0.4, 0.4--0.6, and $>$0.6, respectively. Figure \ref{fig:1D_Z2}
(botton panels) shows that for rescaled galaxies the $S/N$ can be so
low that it even affects the innermost part (i.e., bulge) of the
galaxy. Note that, in order to mimic higher redshift observations, the
range of $S/N$ is different (lower) for rescaled galaxies.  On the
other hand, as a result of more  
homologous light distribution, the fraction of single-component fits
with $n > 6$ is lower for rescaled galaxies with $B/T < 0.4$. 

\subsection{Fitting Disk Galaxies with Two Components}

\subsubsection{Measuring $B/T$ and Total Luminosities}

Although the initial values do not have a major effect
  on the single S\'{e}rsic fits, they require some consideration for
  multiple-component decompositions, where it is generally advisable
  to use an additional, external prior or information. For example,
  one approach might be to fit a single component first, then add a
  second to the result, differentiating the two in some way.  Another
  approach, which we adopt here, is to estimate $B/T$ using
  a one-dimensional light profile. Knowing that the disk component
  follows an exponential profile, we look for the part of the profile
  that traces a straight line in logarithmic space. A straight line
  fitted (in logarithmic space) to that section of the light
  distribution provides an estimate for the disk central surface
  brightness and scale length. Assuming the total brightness is
  given (which can be robustly estimated by single S\'{e}rsic fits),
  the bulge component magnitude and therefore the galaxy $B/T$ can be
  estimated.  Most numerical problems encountered in the fit are
  caused by the small bulge sizes in our study, because they are near
  the resolution limit at high redshifts.  When the sizes go
  below 0.1 pixel during a fit we hold them fixed to 0.5 pixel.  

\paragraph{Local Galaxies}

Figure \ref{fig:BTZ0} shows the results of measuring the total
luminosity of the bulge and disk and $B/T$ for galaxies with three
different methods: bulge with free S\'{e}rsic + exponential disk, two S\'{e}rsic
components, and bulge with $n$=4 + exponential disk.  
For disk-dominated galaxies (left column), $B/T$ and bulge and disk
luminosities can be measured robustly, independent of the method. 
 For galaxies with larger $B/T$, Figure \ref{fig:BTZ0} shows that
 fixing the bulge S\'{e}rsic index to incorrect values can lead to
 considerable biases, especially for galaxies with intermediate $B/T$
 ($0.4 < B/T \leq 0.6$); while the bulge flux contribution is
 significant, its light distribution does not necessary follow a
 de~Vaucouleurs profile. Using this method tends to
   overestimate the bulge total flux and does the
   opposite for the disk total flux. This effect becomes more
   prominent at lower $S/N$. However, fitting model galaxies with a free
 S\'{e}rsic + exponential disk (cyan points) 
and even two S\'{e}rsic components (green points) results in more robust
luminosity and $B/T$ determinations. 

Previous studies have shown that not all disks follow an exponential
light distribution (e.g., \citealt{Boroson81}), and therefore it is
useful to know how well one can measure the S\'{e}rsic index and
subsequent properties of the disk component, in case the disk has $n$
different from 1. Comparing the cases where the disk is set to $n = 1$
versus $n$ = free in Figure \ref{fig:BTZ0}, we find that although in
general the uncertainties are lower for $n = 1$, at higher $S/N$ the
results of $n = 1$ and $n$ = free are comparable. This implies that,
as known, the disk component need not follow a pure exponential
function, in oder to find a reliable bulge+disk model fit.  

\paragraph{Rescaled Galaxies to \emph{z} = 2 }

As expected, the measurements are more uncertain for high-\emph{z}
galaxies (Fig. \ref{fig:BTZ2}), as their bulges typically have $R_e$
smaller than one pixel, and, besides, structural non-homologies are
mostly washed out (Fig. \ref{fig:1D_Z2}).  However, by employing
different fitting methods, $B/T$ can be measured with no significant
bias for $S/N$ of a few 100, values typical of actual {\it HST}\
observations of compact massive galaxies (e.g., \citealt{Szomoru12}).
Similar  to the result for local galaxies, for galaxies with very
small $B/T$, any of the methods yields reliable results; however, for
galaxies with higher $B/T$, it is best to fit the bulge component with
a free $n$ to mitigate systematic biases, even if the biases are
smaller compared to local galaxies.  

\subsubsection{Measuring Bulge Properties}

\paragraph{Local Galaxies}

Figure \ref{fig:bulgeZ0} reinforces the conclusions drawn from Figure
\ref{fig:BTZ0}. S\'{e}rsic + exponential and S\'{e}rsic + S\'{e}rsic
models can measure bulge properties with high accuracy. Figure
\ref{fig:bulgeZ0} shows that $n=4$ + exponential is not a reliable
model as it leads to biases in bulge size measurements. Fixing bulge
$n$ to 4 tends to overestimate its effective radius. More importantly,
any information on bulge $n$ is lost. 

\paragraph{Galaxies Rescaled to \emph{z} = 2 }

Figure \ref{fig:bulgeZ2} shows that for high-\emph{z} galaxies with
$B/T \leq 0.4$, measuring the bulge properties is vulnerable to large 
uncertainties and systematic offsets independent of the fitting
method. For our model galaxies with $B/T$ $\leq$ 0.2, the
  systematic errors on the bulge properties measurements are as large
  as 50\%, even when $S/N$ is high, due mainly to the small angular resolution. 
However, for model galaxies with 0.2 $\le B/T \leq$ 0.4 and very high
$S/N$ values, the size and ellipticity can be measured with little error.
Note that the bulge effective radius of these low-$B/T$ galaxies are $\approx$ 0.5
  pixel (or less) and thus mainly unresolved. However, for model
galaxies with $B/T \geq 0.4$ and at $S/N$ of compact massive galaxies
observed  at high \emph{z}, the bulge properties can be measured
reliably. As expected, the S\'{e}rsic index is most vulnerable to large
uncertainties. Similar to the results of local galaxies, the errors are greater for $n=4$ + exponential fits. 

The high levels of uncertainties in bulge ellipticity measurements may
be surprising. For idealized model galaxies, like ours, the
ellipticity is expected to be one of the more robust
measures. The robustness of recovered ellipticity is directly related to the recovered size.
At lower redshifts, there is hardly any uncertainties but at higher redshifts,
especially at lower $S/N$ and for galaxies with lower $B/T$, the structural
parameters are very uncertain as the bulge $R_e$ cannot be resolved by
{\tt GALFIT}. The recovered bulge and disk ellipticity distributions are
biased toward extreme values. This is partly also an artificial effect of
having hard numerical boundaries for the ellipticities between 0 and 1
rather than being continuous between arbitrarily negative and positive values.  When
nearly all the flux fits inside a single pixel the axis ratio
parameter becomes irrelevant. 

The uncertainties for measuring bulge structural
  properties are higher than for measuring the total magnitude and
  $B/T$. This is because for a given overall profile shape (total flux),
  component parameters will have covariances consistent with that
  shape, to within the noise (often thought of, inaccurately, as
 ``degeneracy''). This is especially true here because most high-\emph{z}
  bulges fit inside a single pixel; naturally there is considerable
  uncertainty in the sizes and ellipticities but less so the
  luminosity.” 

\subsubsection{Measuring Disk Properties}

The number of pixels that each component occupies affects the
robustness of the fit.  The disk component, being intrinsically larger
than the bulge, is therefore easier to measure than the bulge. Figure
\ref{fig:diskZ0} demonstrates this fact. All methods can measure the
disk properties of local galaxies with no systematic offsets over a
wide range of $S/N$.  The only exception is when the $n$=4+exponential
model is used for galaxies with higher $B/T$; this method causes
$\sim$10\% systematic offsets.  The middle panel of the figure shows
that the S\'{e}rsic+S\'{e}rsic model can measure the disk S\'{e}rsic
index robustly. 

Figure \ref{fig:diskZ2} shows that even for galaxies rescaled to
\emph{z} = 2, disk properties can be measured with little to no
systematic offsets and with low uncertainties at the $S/N$ pertinent
to massive compact high-\emph{z} galaxies. Measuring
  the disk S\'{e}rsic index has the highest level of
  uncertainties. Although the random uncertainties are near zero at
  higher $S/N$, they increase to 10-20\% at intermediate $S/N$, and
  larger at very low $S/N$. Note that the measured disk $n$ is centered
  on 1 and therefore there is no systematic bias.

\section{Comparison to Other Studies}

Our results generally agree well with those of other similar recent
studies.  Using a sample of galaxies spectroscopically selected from
SDSS, \citet{Meert13} show that single-component
S\'{e}rsic fits of two-component (bulge + disk) simulated galaxies can
lead to systematic biases in size measurements (their Fig. 8b), but
the offsets are small, generally \lax 10\% over a wide range of galaxy
sizes.  This is consistent with our simulations of local galaxies:
Tables 1 and 2 show that for our disk-dominated and bulge-dominated
model galaxies the systematic offsets are less than 10$\%$, and for
model galaxies with more prominent non-homology the offsets can be as
large as 20$\%$. However, for very low $S/N$ (i.e., a few hundred) the
systematic offsets can be even larger. 

\citet{Meert13}, again in agreement with our findings, show that
fitting local bulge+disk model galaxies with two components result in
more reliable fits and measured structural parameters (their Figs. 9
and 10), as the systematic offsets are close to zero over a wide range
of apparent magnitude (i.e., $S/N$). They also find that disk
parameters can be measured more accurately compare to those for the
bulge (Fig. \ref{fig:diskZ0}). The parameter most vulnerable to large
uncertainties is $n_{b}$, confirming our findings
(Fig. \ref{fig:bulgeZ0}).   

\citet{Meert13} further examined the effects of image size on the
accuracy of {\tt GALFIT} sky determination. They find (their Figs. 12,
14, and 15), as we do, that {\tt GALFIT} can measure the sky value
with an accuracy of 0.1$\%$. Our simulations indicate that when the
images are large enough (i.e., 10--15 times larger than effective
radius of the galaxy), the sky component can be set as a free
parameter for {\tt GALFIT} modeling. 

The simulations of \citet{Mosleh13} show that size measurements done by two-component S\'{e}rsic
fits are more reliable than single S\'{e}rsic fits. They found
that for local massive, red, early-type galaxies (their Fig. 3),
single-component fits overestimate the size. This bias goes away for
their redshifted galaxies, consistent with our
findings. Rescaling galaxies to higher redshifts washes out the
structural non-homology and therefore single S\'{e}rsic fits can model
the galaxy light profile better.

Similar, but less general sets of simulations are
  performed by Bruce et al. (2014). Their simulated galaxies are
  $n_{\mathrm{bulge}}$ = 4 + $n_{\mathrm{disk}}$ = 1 models at
discrete $B/T$ and effective radii values ranging from 0.01 to 0.99,
and 1 to 20 pixels, respectively. All their mock galaxies have the
same brightness, and thus the effect of $S/N$ is not examined. For
their simulated galaxies with 0.1 $< B/T <$ 0.9, they find that the
derived sizes of the bulge component have higher uncertainties than
the disk components, which our findings confim. Bruce et
al. concluded that $B/T$ can be measured to within 10\% accuracy. The
simulations of Lang et al. (2014) show similar measurement errors in
$B/T$ for their \emph{z} $\approx$ 2 modeled galaxies. Our simulations
further show that the $B/T$ measurement error depends on the galaxy
$S/N$, and it ranges from only a few percent at higher $S/N$ to 10-20\%
at lower redshifts.

In general, our results confirm the findings of similar previous studies. Furthermore, our simulations, for
the first time, provide a detailed study of the $S/N$ effect on the {\tt GALFIT} modeling of
galaxies with a wide range of $B/T$. Our
results show that for different methods (i.e. S\'{e}rsic+exponential,
$n$=4+exponential), the reliability of the fits varies for different
values of $B/T$. Knowing the optimum method for a specific value of
$S/N$ and $B/T$ offers a roadmap for a variety of structural
analysis. 

\section{Implications for Red Nuggets}

Our simulations provide a metric for quantifying the uncertainties in
measuring the properties of the red, massive galaxies at \emph{z}
= 2. These peculiar objects are found to be generally compact ($R_e$
$\leq$ 2.0 kpc). They mainly have $H_{160}$ $<$ 23, with a small fraction having
$H_{160} \approx 23-24$  (\citealt{Szomoru12}). This translates to
high $S/N$ ($\geq$ 100) for typical single-orbit {\it HST}\ exposures. 

Several previous studies have shown that if the light
distributions of red nuggets follow a single S\'{e}rsic profile, at the $S/N$ that
these galaxies are observed, the structural parameters and the total
luminosity can be measured accurately  (e.g., \citealt{Haussler07};
\citealt{Trujillo07}; \citealt{Cimatti08}; \citealt{Szomoru10};
\citealt{vanDokkum10}; \citealt{Williams10}; \citealt{Papovich12}; 
\citealt{vanderWel12}; \citealt{Meert13}; \citealt{Mosleh13}; Davari
et al. 2014). 

On the other hand, the structural non-homologies can lead to biases in
size measurements using single-component fitting (a popular
method). Davari et al. (2014) found that if the red nuggets have
structures similar to those of local massive elliptical galaxies, single
S\'{e}rsic fits return reliable size (with about 10$\%$ systematic
offsets) and total luminosity measurements. In this paper, we further
extend previous work by examining the potential biases of
single-component size measurements of red nuggets, {\it assuming}\
that their light distributions resemble that of local galaxies with a
bulge {\it and}\ a disk component.  Our work is motivated by the possibility that red nuggets
contain a significant disk component.  We are interested in knowing
whether the presence of a disk affects the overall size and luminosity
measurements of red nuggets, as well as the prospects of decomposing
the bulge and disk components, to study derive their structural
parameters and eventually their redshift evolution. 

Figure \ref{fig:1comp_Z2} and Tables 3 and 4 show the best-fit results of 
the single-component S\'{e}rsic fits of the bulge+disk galaxies.  With
adequate $S/N$ ($\geq$ 100, comparable to red nuggets studied in
CANDELS), the presence of a disk has an insignificant effect (a few percent underestimation)
on their sizes. This holds over a wide range of sizes ($1 <
R_e < 25$ pixels). At much lower $S/N$ values (e.g., $\leq$ 50),
the systematic offsets are about 20$\%$, which is still
insignificant compared to the estimated amount of evolution (200\%--500\%).   Conclusion: if red
nuggets have structures similar to those of local spiral galaxies,
single-component S\'{e}rsic fits of these galaxies measure their global sizes
robustly. Furthermore, Table 4 shows that the total luminosity of the
bulge+disk galaxies can be measured very accurately, too.

The S\'{e}rsic indices of the best-fit single-component models can be an
indicator of the $B/T$ of the galaxy (\citealt{Fisher08}). We found that, at
the $S/N$ of observed red nuggets, all the best-fit models with $n
\leq 2$ have $B/T < 0.3$. It is more difficult to physically interpret
fits with $n \approx 4$, as it can indicate either 
a high $B/T$ or a galaxy with moderate $B/T$ (e.g.,
$\sim$0.4) and moderately low $S/N$ (Fig. 
\ref{fig:1D_Z0}, middle panels). Lower values of S\'{e}rsic indices are
more informative and less difficult to interpret. 

We also examine how well one can measure the $B/T$ of disk galaxies
and the reliability of the resulting parameters for the bulge and disk components.
Figure \ref{fig:BTZ2} shows that for galaxies with $B/T
\geq 0.2$ and at the $S/N$ pertinent to red nuggets, $B/T$ can be
accurately measured from fitting a S\'{e}rsic + exponential
disk model. 
 
The properties of bulges can be measured
robustly in galaxies with $B/T \geq 0.4$ and with
little to no systematic offsets but with large uncertainties
in those with $0.2 \leq B/T < 0.4$ (Fig. 
\ref{fig:bulgeZ2}). The best strategy is to leave the S\'{e}rsic index
of the bulge free, while adding an exponential component for the disk.
Fixing the bulge S\'{e}rsic index to $n = 4$ introduces systematic
biases, apart from removing any information on $n$.
At higher $B/T$, especially, using this method tends to
  overestimate bulge total flux and effective radius and underestimate
  those of the disk component. Our simulations show
  that recovering the bulge properties of low $B/T$ at \emph{z} $\approx$
  2 is challenging due to present resolution level of {\it HST} WFC3.
  For well-resolved bulges, even at very low $B/T$, the bulge
  properties can be measured well, as our low-redshift simulations
  show. 

The properties of the disk are easiest to measure and least subject to
serious systematic error (Fig. \ref{fig:diskZ2}).  This is
expected, as the disk component is the most extended and is least
effected by the PSF.  As for the case of the bulge mentioned above,
the best strategy is to set the bulge  S\'{e}rsic index free
while fixing the disk to an exponential.  Of course, this imposes the
assumption that the disk follows an exponential light distribution. 

The analysis of massive galaxies at \emph{z} $\approx$
  2 find bulge sizes that are larger than those assumed in our simulations (Bruce et al. 2014). The
  fact that they are not in our sample, comprised only of scaled
  versions of local bulge+disk systems, suggests that the actual
  population of galaxies at $z \approx 2$ differs from that our
  assumed $z \approx 0$ sample. The fraction of galaxies with $B/T$ $<$ 0.4 among the real galaxies
  at high-\emph{z} is relatively low, which further confirms that we are
  comparing two different populations.  Our simulations indicate that
  the actual galaxy population at high redshifts is easier to analyze compared with
  the unresolved bulge cases because they span an intermediate angular
  size between low-\emph{z} and \emph{z} $\approx$ 2 galaxies. 

\section{Summary}

A considerable fraction of the compact, red massive galaxies at \emph{z} $\approx$ 2 may have a disk component. This motivated us to simulate mock observations of model galaxies to investigate the extent to which a disk component, if present, can be detected under realistic conditions typical of actual observations. 

The simulated bulge+disk galaxies span a uniform and wide range of $B/T$, 
and we constrain the disk and bulge components to follow empirical
scaling relations established for local galaxies.  We then rescale these $z \approx 0$ galaxies to
mimic the $S/N$ and sizes of galaxies observed at \emph{z} $\approx$ 2. This
provides the most complete set of simulations that allows us to examine the
robustness of two-component image decomposition of compact disk galaxies at
different $B/T$. 

First, we measure the basic structural parameters using single
S\'{e}rsic fits, with special emphasis on their sizes. This analysis method is
popularly employed in the literature. We then study the
robustness of different methods of bulge+disk decomposition of these
composite galaxies. Furthermore, we assess the effectiveness of different sky background
fitting methods. 

For the range of $S/N$ and sizes pertinent to red nuggets, we conclude: 

\begin{itemize}

\item Modeling bulge+disk galaxies with a single S\'{e}rsic component
  does not bias the sizes too low, by no more than 
  10$\%$. The apparent compactness of red nuggets is real; it is
  not the result of missing faint, outer light. However,
  single-component fits of galaxies with low $B/T$ preferentially
  leads to relatively larger biases in size and luminosity.

\item The $B/T$ can be measured accurately, regardless of the $B/T$. 

\item Bulge properties of galaxies with $B/T$ \gax\ 0.4 can be
  measured robustly.  This becomes increasingly difficult for galaxies with  $B/T$ below this limit.

\item Disk properties are subject to the least amount
  of systematic and random error, regardless of the $B/T$. 

\item Fits with S\'ersic indices larger than 6 have larger
  uncertainties and can cause significant systematic errors. Refitting
  these galaxies by fixing $n$ to 6 provides
  more reliable results.  

\item {\tt GALFIT} can measure the sky value accurately when enough
  background pixels are available at high $S/N$.  At low $S/N$, fixing the sky value during the fitting
  reduces systematic errors.

\end{itemize}

\acknowledgements
We thank the anonymous referee for a careful reading of the paper and
for making detailed suggestions that improved the understanding of this work.
RD has been funded by a graduate student fellowship awarded by Carnegie Observatories; we are grateful to Wendy Freedman for her support.   LCH acknowledges support by the Chinese Academy of Science through grant No. XDB09030102 (Emergence of Cosmological Structures) from the Strategic Priority Research Program and by the National Natural Science Foundation of China through grant No. 11473002.  RD thanks Gabriela Canalizo and Heather L. Worthington for providing long-term support and Andrew Benson and Shannon Patel for useful discussions.

\clearpage


\end{document}